\documentclass[manuscript]{aastex61}

\begin{document}

\title{Stellar Parameters and Radial Velocities of Hot Stars in the Carina Nebula}

\author{Richard J. Hanes}
\affil{Department of Physics, Lehigh University, 16 Memorial Drive East, Bethlehem, PA 18015, USA}
\email{rjh314@lehigh.edu}

\author{M. Virginia McSwain}
\affil{Department of Physics, Lehigh University, 16 Memorial Drive East, Bethlehem, PA 18015, USA}
\email{mcswain@lehigh.edu}

\author{Matthew S. Povich}
\affil{Department of Physics \& Astronomy, California State Polytechnic University Polytechnic, Pomona, CA 91768, USA}

\begin{abstract}

The Carina Nebula is an active star forming region in the southern sky that is of particular interest due to the presence of a large number of massive stars in a wide array of evolutionary stages. Here we present the results of the spectroscopic analysis of 82 B-type stars and 33 O-type stars that were observed in 2013 and 2014.  For 82 B-type stars without line blending, we fit model spectra from the Tlusty BSTAR2006 grid to the observed profiles of H$\gamma$ and He $\lambda\lambda$ 4026, 4388, and 4471 to measure the effective temperatures, surface gravities, and projected rotational velocities. We also measure the masses, ages, radii, bolometric luminosities, and distances of these stars. From the radial velocities measured in our sample, we find 31 single lined spectroscopic binary candidates. We find a high dispersion of radial velocities among our sample stars, and we argue that the Carina Nebula stellar population has not yet relaxed and become virialized. 

\end{abstract}


\section{Introduction} \label{sec:intro}

The Carina Nebula is one of the most active star forming regions located nearby in our Galaxy, containing many massive stars spanning across the evolutionary spectrum. The brightness, proximity, and size of Carina (more than 1 deg$^2$ on the sky) make it an ideal candidate for study as it provides a window into the entire stellar formation and evolution process. It has been the subject of several recent, large surveys including the Chandra Carina Complex Project (CCCP) \citep{Townsley2011}, the VISTA Carina Nebula Survey \citep{Preibisch2014}, and the VST Photometric H$\alpha$ Survey (VPHAS+) \citep{Mohr2017}.    \\

With over 200 massive OB stars \citep{Gange2011} and 1400 young stellar objects (YSOs; \citealt{Povich2011a}), study of physical parameters of these stars can provide insight into stellar formation across the nebula. \cite{Huang2006} measured the effective temperature ($T_{\rm eff}$), surface gravity ($\log g$), projected rotational velocity ($V \sin i$), and helium abundance of 39 stars spread across clusters Collinder (Coll) 228, Trumpler (Tr) 14, and Tr 16 in Carina. \citet{Berlanas2017} recently performed a preliminary study of 14 O-type stars observed in the Gaia-ESO Survey (GES).  While both of these studies give an initial look into the spectroscopic parameters of the stars in Carina, there is still a need for a broader study of the stars throughout the rest of the nebula. \\

In a previous paper \citep{Alexander2016}, we spectroscopically classified 36 O-type stars and 128 B-type stars scattered throughout the nebula, confirming 23 new OB-type stars. We present here the results of the measurements of physical parameters of the observed B stars from \cite{Alexander2016}. This paper should provide the current largest and most comprehensive catalog of spectroscopic parameters of massive stars in the Carina Nebula.\\

 Section 2 briefly describes the observations and data reduction of the spectra. In Section 3, we discuss how we measured, via model fitting with the Tlusty BSTAR2006 grid, $T_{\rm eff}$, $\log g$, and $V \sin i$ of these stars. Comparing these results to the evolutionary tracks and isochrones, we also measure the mass, radius, and age. We also compare our results with any shared stars in past studies. Section 4 discusses the radial velocities and distances of the stars in our sample.  \\

\section{Observations} \label{sec:obs}

Observations of the stars were made at the Anglo-Australian Telescope (AAT) over the course of two runs in March 2013 and April 2014. The observations of these stars are described in greater detail by \cite{Alexander2016}.  We chose two different wavelength regions, 3925-4210\AA~(2013) and 4235-4510 \AA~(2014), to cover many useful H and He lines for analysis.   As the target spectra were vertically stacked on the imaging plane, distortions in the imaging plane meant that the exact spectral coverage varied among the targets and sky spectra.  During our first day of observations in 2014, we used a slightly different range (4200-4475 \AA) for some of our exposures, but we found that this omitted the He I $\lambda$4471 line for some of our targets due to variable dispersion across the chip. \\

The raw spectra were reduced using the \textsc{dohydra} package of IRAF and a custom IDL code for sky subtraction to account for the changing wavelength coverage across the CCD.  Due to the variable dusty nature of the Carina Nebula, sometimes the average sky spectrum is too strong or too weak in comparison to our targets, which results in contamination of the Balmer line cores for some of our stars.  \\

Bright stars in our 2013 data generally have a signal-to-noise ratio (S/N) of 50-120, while the faint stars have a S/N of 30-70. The bright stars in our 2014 data have a S/N of 100-200, while the faint stars have a S/N of 120-210. The signal-to-noise of our 2013 data was low because our observing time was cut short due to wildfires in the area. We used two different fiber configurations for the bright versus the faint stars and observed them with different exposure times, which is how we achieved marginally better S/N for the fainter stars. Our measurements of S/N for each star are listed in Table \ref{tab:vsini}.\\

\section{Stellar Physical Parameters} \label{sec:spp}

We used the non-local thermodynamic equilibrium (NLTE) Tlusty BSTAR2006 \citep{Lanz2007} model spectra to measure $T_{\rm eff}$, $\log g$, and $V \sin i$ of our observed B-type stars. BSTAR2006 offers several grids with different metallicities and microturbulent velocities. For our purposes, we assumed a solar metallicity ($Z/Z_\odot$~=~1) and a microturbulent velocity of $V_t~=~2$ km s$^{-1}$. The value for microturbulent velocity is not very important in this situation because He I $\lambda\lambda$ 4471, 4388, and 4026, which we used to measure $V \sin i$, are not very sensitive to $V_t$ \citep{Lyubimkov2004}.\\

Before fitting the stars, we first estimated the $T_{\rm eff}$ and $\log g$ for a star based on the strength and shape of the Balmer and helium lines in our spectra. Using custom IDL codes, we measure $V \sin i$ by artificially broadening the model spectra for instrumental and rotational broadening across a series of 10 km s$^{-1}$ steps. We then compare the sum of the squares of the residuals ($\Sigma$(O-C)$^2$) for each step and determine minimal value of a parabolic fit as the value for V sin $i$ for a given line. The error associated with this measurement was calculated by finding fits that fell at or below a 5\% tolerance in $\Sigma$(O-C)$^2$. We used the He I $\lambda\lambda$ 4026, 4388, and 4471 lines for the fitting process, and then a weighted average is calculated as our measured V sin $i$. The measurements for $V \sin i$ of all the helium lines as well as the weighted average are recorded in Table \ref{tab:vsini}. \\

After measuring V sin $i$, we then modeled the spectra at H$\gamma$ for $T_{\rm eff}$ and $\log g$ along each point in the BSTAR2006 grid.  Once we found the closest fits within grid, we then interpolated between grid points via a linear scaling of the models to find the best fit for $T_{\rm eff}$ and $\log g$.  Errors for $T_{\rm eff}$ and $\log g$ were calculated by finding fits that fall at or below a 10\% tolerance in the $\Sigma$(O-C)$^2$ because we fit for the two parameters simultaneously. $T_{\rm eff}$ and $\log g$ are recorded in Table \ref{tab:Teff}. \\

The earliest B stars have temperatures near $T_{\rm eff}$ = 30,000 K, at the edge of the BSTAR2006 grid, resulting in errors in $T_{\rm eff}$ that are likely underestimated. Comparing our measured $T_{\rm eff}$ with the spectral types in \citet{Alexander2016}, we note that some of the stars appear hotter than expected. A hotter B-type star has a smaller H$\gamma$ equivalent width, and their expanding atmospheres will produce emission that will partially fill in the line profile further, making the star appear hotter still. These stars will be analyzed again along with the O-type stars in our sample in a future paper. On the other hand, other B stars appear cooler than expected, which may be a result of unseen binary line blending. Here, the cooler companion's H$\gamma$ profile is artificially increasing the measured line strength. Overall, we find that the temperatures of our earliest B-type stars may be up to 9,000 K cooler than their spectral type suggests, so our formal errors represent only part of the true uncertainty.\\  

Due to the difficulty of sky subtraction and the brightness of Carina in the hydrogen recombination lines, our spectra frequently have nebular contamination at the cores of the Balmer lines that are challenging to properly account for. When present, we ignore the affected wavelengths during the fitting process. This leads to larger errors in our measurements for $T_{\rm eff}$ because we lose information about the line core. \\

\begin{figure}
\plotone{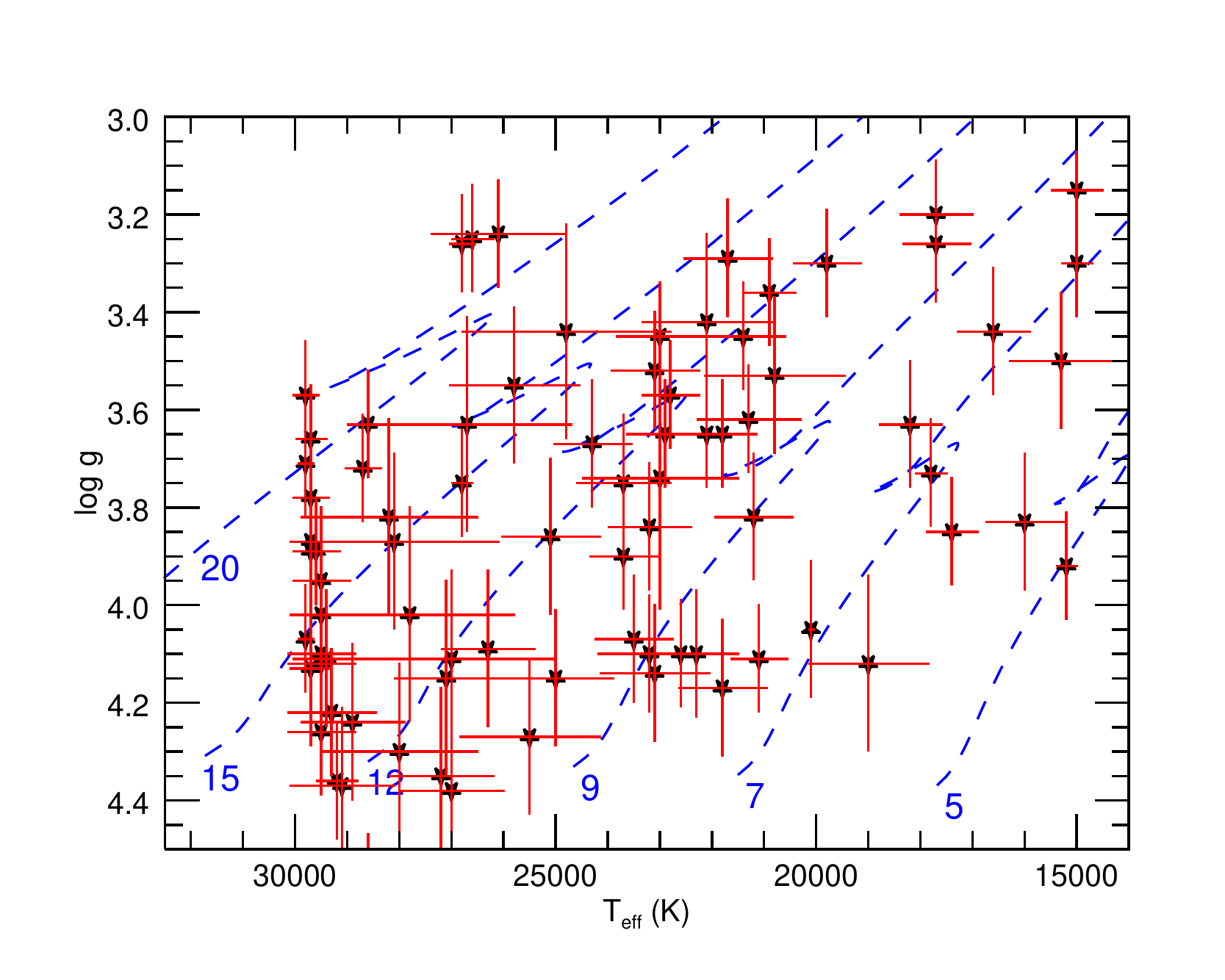}
\caption{HR-Diagram of our observed B-type stars using the evolutionary tracks offered by \citet{Ekstrom2012}. Stars near $T_{\rm eff}$ = 30,000 K are at the edge of the BSTAR2006 grid and their errors in $T_{\rm eff}$ are likely underestimated. The values associated with each evolutionary track are in solar masses.}
\label{fig:HR}
\end{figure}

As the effect of $\log g$ is mostly present in the wings of the spectral lines, we find that continuum fitting causes a systematic error of $\Delta \log g_{syst} \sim 0.1$ dex.  To compensate for this systematic error,  the errors, $\Delta \log g$, presented in Table \ref{tab:Teff} are computed using the formal error from our model fitting added in quadrature with the systematic error from the continuum fitting process. We also include an HR-Diagram of our observed stars in Figure \ref{fig:HR}.\\

\begin{figure}
\gridline{\fig{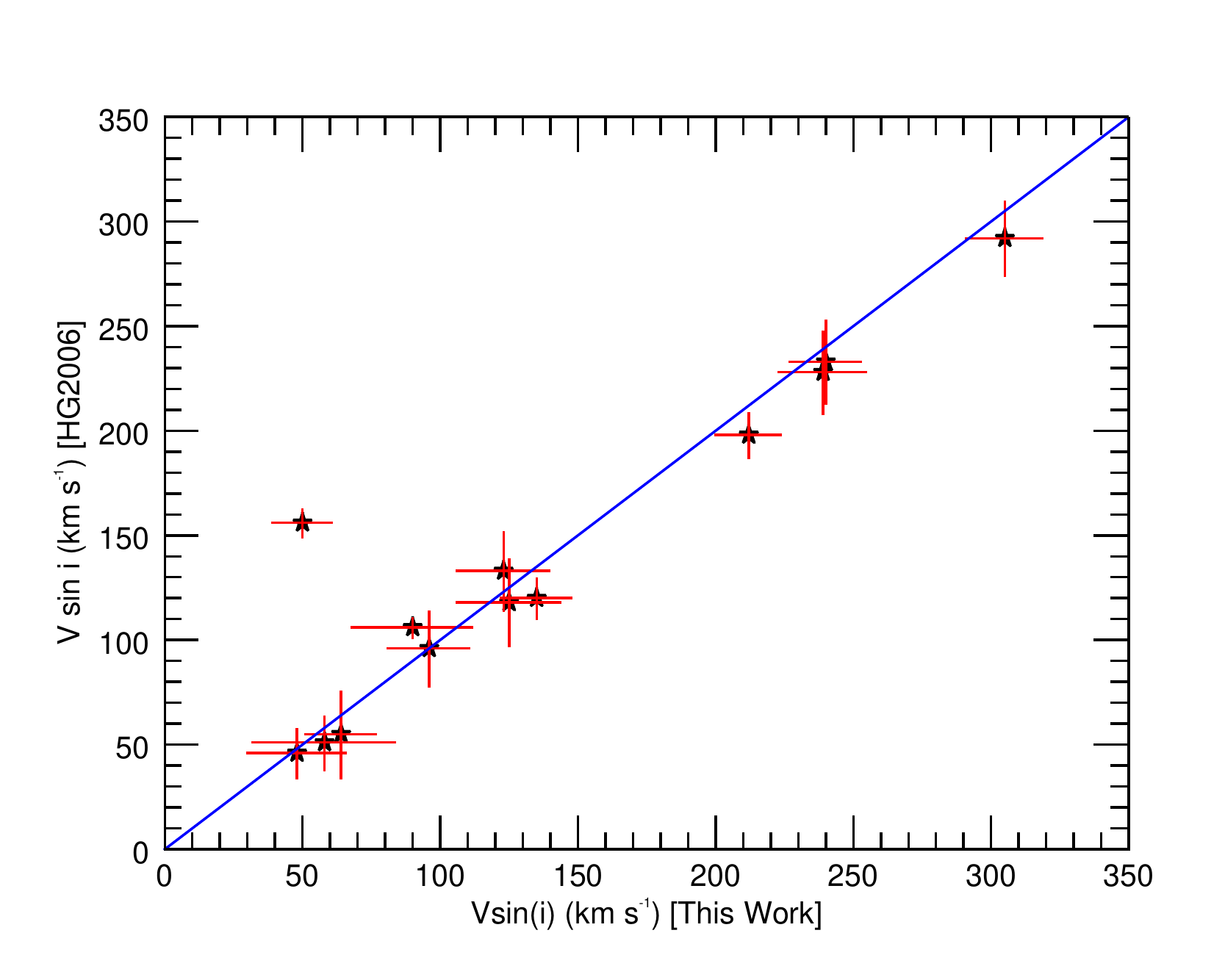}{0.40\textwidth}{(a)}
          \fig{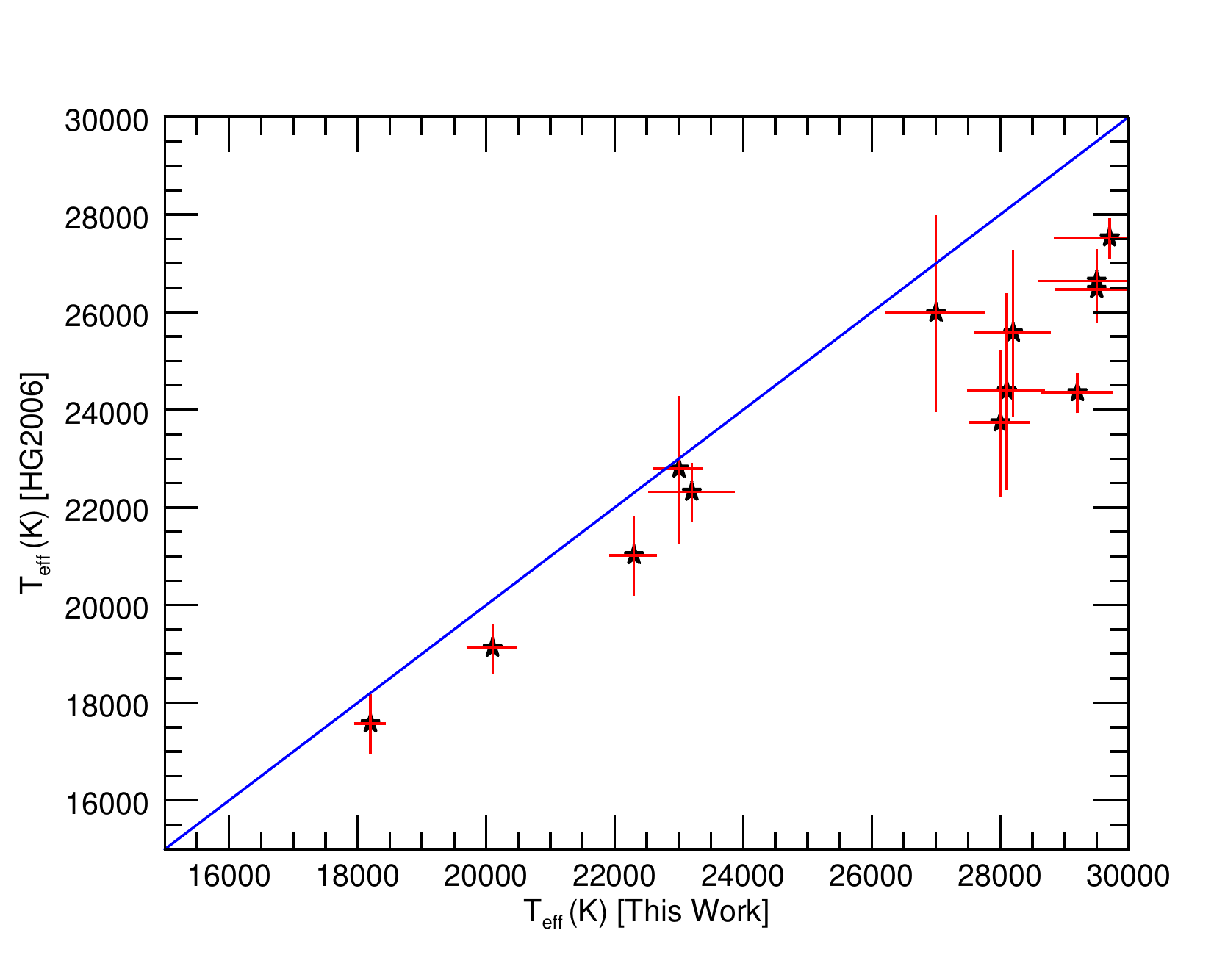}{0.40\textwidth}{(b)}}
\gridline{\fig{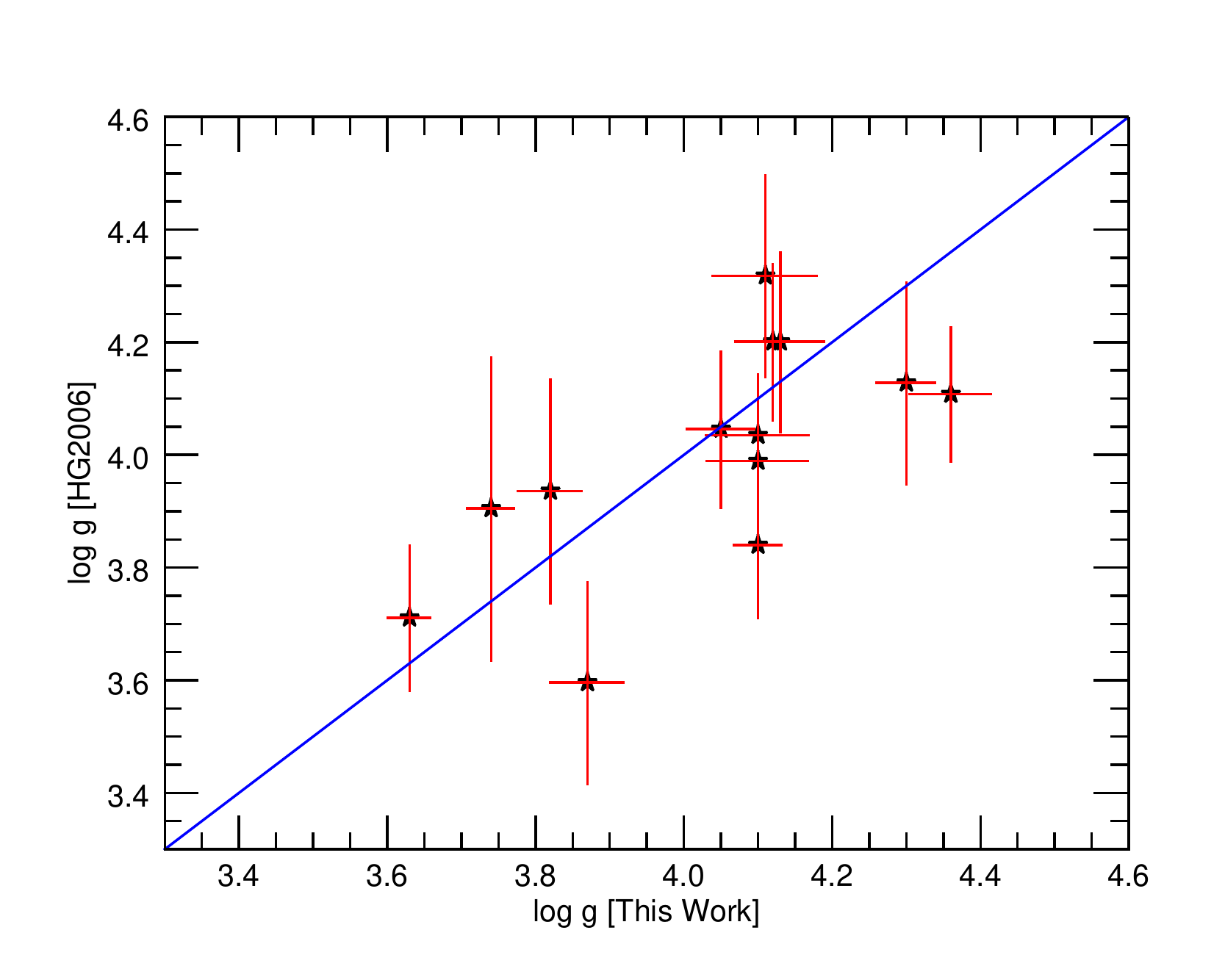}{0.40\textwidth}{(c)}}
\caption{Comparison of our results with \citet{Huang2006}. The horizontal and vertical error bars in these figures are the calculated errors from this work and \citet{Huang2006}, respectively. }
\label{fig:HvH}
\end{figure}

\citet{Huang2006} performed a similar analysis among a sample of O- and B-type stars in Carina, of which we share 13 B stars. A comparison of our results is shown in Figure \ref{fig:HvH}.  Our $V \sin i$ measurements agree within error, with the sole exception of Tr 16-25. We suggest that Tr 16-25 may be a double-lined spectroscopic binary (SB2) considering the large discrepancy between our results. It's possible that the different observation times could have caught the binary system at different stages of the orbit, resulting in different blends of the spectral lines, so measurements of $V \sin i$ would differ. \\

The effective temperatures that we measured seem to have a noticeable increasing trend as the temperature increases. In their paper, \cite{Huang2006} use the local thermodynamic equilibrium (LTE) ATLAS9 atmospheric models to measure $T_{\rm eff}$ and $\log g$ in contrast to the NLTE Tlusty models we used. It has been shown that LTE models can sufficiently describe cooler stars below $T_{\rm eff} < $ 22,000 K \citep{Przybilla2011} and that NLTE models are required for hotter O- and B-type stars. This explains the large discrepancy for our stars with $T_{\rm eff} >$ 27,000 K. Our measurements for $\log g$ are consistent with \cite{Huang2006}.\\

A straightforward application of our measurements is to compare $T_{\rm eff}$ and $\log g$ to model evolutionary tracks to measure the mass ($M_\star$), radius ($R_\star$), and age ($\tau_\star$) of the stars in our sample. Using the non-rotating versions of the evolutionary tracks offered by \citet{Ekstrom2012}, we can measure these parameters by doing a linear interpolation between the evolutionary tracks.   These values, as well as the calculated bolometric luminosity ($L_{bol}$), are included in Table \ref{tab:Teff}. The error bars on each quantity are measured by varying $T_{\rm eff}$ and $\log g$ by their respective errors. \\

\begin{figure}
\centering
\includegraphics{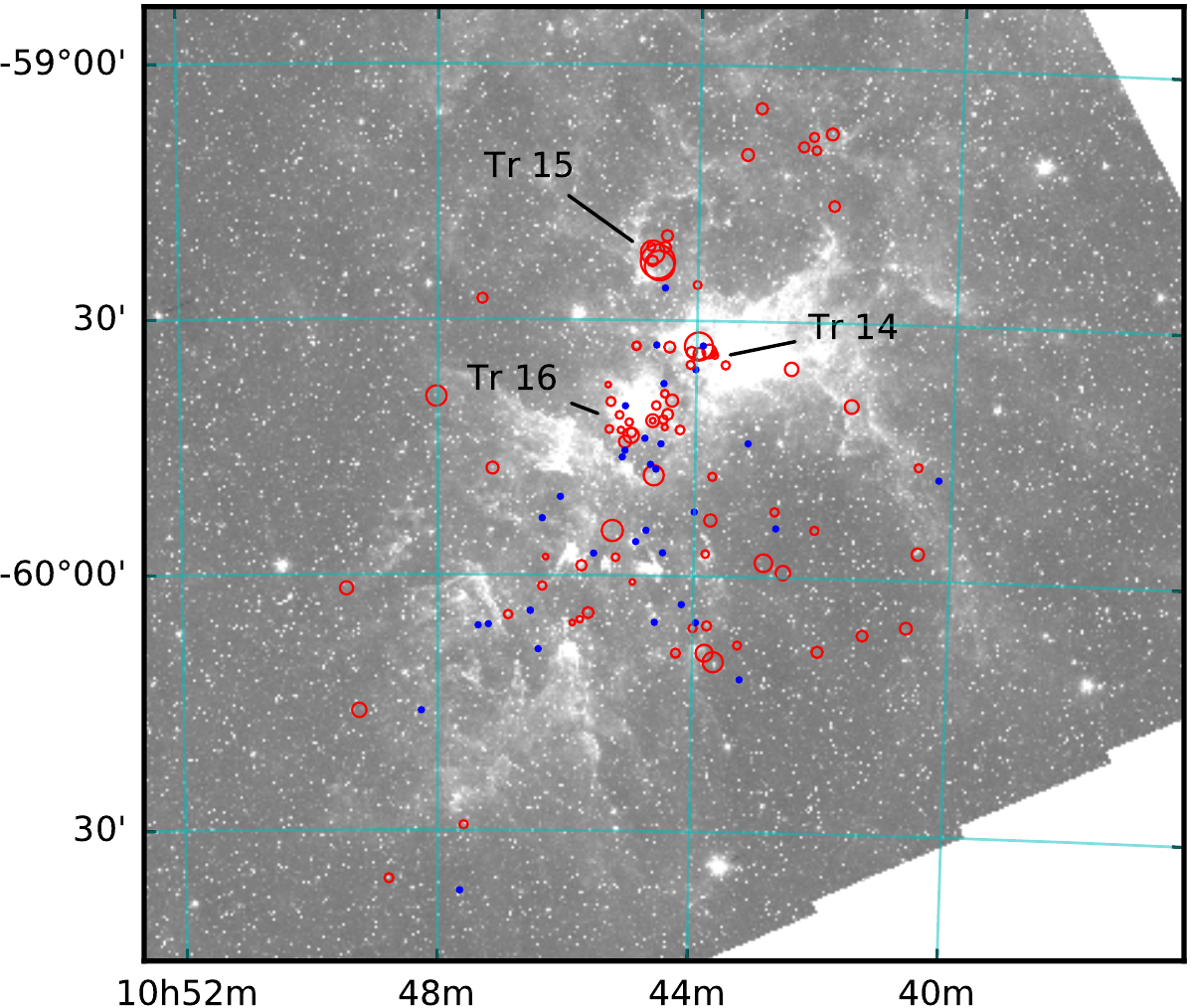}
\caption{Age distribution of the B stars (red) across Carina. The sizes of the bubbles are relative to the magnitude of the age of the stars. The O stars from our sample are also included in blue. The positions of the clusters Tr 14, Tr 15, and Tr 16 are marked. }
\label{fig:Age_Pos}
\end{figure}

As an active star-forming region, the distribution of the stars across the nebula can provide insight into the structure and features of Carina. The stellar age distribution in Figure \ref{fig:Age_Pos} shows that while there are very young stars scattered throughout the nebula, the oldest B stars in the nebula reside in the Tr 15 cluster. The overall dearth of O-type stars in that cluster, as well as the notably thinner nebulosity in the Tr 15 region, suggest that Tr 15 is the oldest cluster in the nebula. Our results also indicate that Tr 14 is a younger cluster than Tr 16, agreeing with similar conclusions in \citet{Damiani2017} and the WEBDA database.  \\

\section{Radial Velocities} \label{sec:RV}

The public data releases from the Gaia-ESO Survey (GES) have been highly anticipated for their calibrated spectroscopy and derived astrophysical parameters. With this in mind, we measured the radial velocities ($V_r$) of all observed O and B stars using a simple parabolic or Gaussian fit of the core of the more prominent spectral lines. We used 12 spectral lines \footnote{The lines used for this were He I $\lambda \lambda$ 4009, 4026, 4120, 4144, 4388, and 4471, Si IV $\lambda \lambda$ 4089 and 4116, C II $\lambda$ 4267, Mg II $\lambda$ 4481, H$\delta$, and H$\gamma$.} across both wavelength regions. Comparing the radial velocities across both epochs, we find that 21 of the 75  (28\%) B stars had $V_{\rm r}$ shifts more than three times the error of the weighted mean $V_{\rm r}$. We classify these as single-line spectroscopic binary candidates (SB1c). The results of our radial velocity measurements of the B-type stars can be found in Table \ref{tab:Vr}.\\

We also include Table \ref{tab:Vr_O} which lists the radial velocity measurements of the O-type stars in our sample. These stars are not the intended focus of this publication, but we believe these results will be beneficial for anyone using the GES data releases for analysis of the Carina region. We found 10 of the 31 (32.3\%) O stars are SB1 candidates. Further work on the physical parameters of these O-type stars will be forthcoming in a future paper.\\

To compare the frequency of our detection of SB1 candidates with what we would expect with a year between measurements, we created a simple code to model the radial velocities of O- and B-type stars. We started by choosing random partners and periods for the stars using the IMF indices calculated by \citet{Kiminki2012}.  The inclination angle and phase angle of the orbit were also randomly generated. We assumed a binary fraction of 30\% $<$ B.F. $<$ 60\% following the study by \citet{Kiminki2012} for binaries with P $<$ 1000 days. We found that we should expect that 19.3 - 38.6\% of O- and B-type stars would exhibit large $V_r$ shifts using only two observations with a year between them. This is consistent with our finding that 29.2\% (31 of 106) of our observed O- and B- type stars are SB1 candidates.  \\ 

In Table \ref{tab:SB2}, we include the radial velocity measurements of the known SB2 systems that we observed that had separate line cores and were not entirely blended together. The radial velocities of the assumed primary star are marked as $V_{r,p}$ while the secondary stars are marked as $V_{r,s}$. The components in HD 303313, HD 93506, LS 1840, and HD 92607 have similar spectral types, making the distinction between the primary and secondary stars impossible. \\

\begin{figure}
\centering
\includegraphics{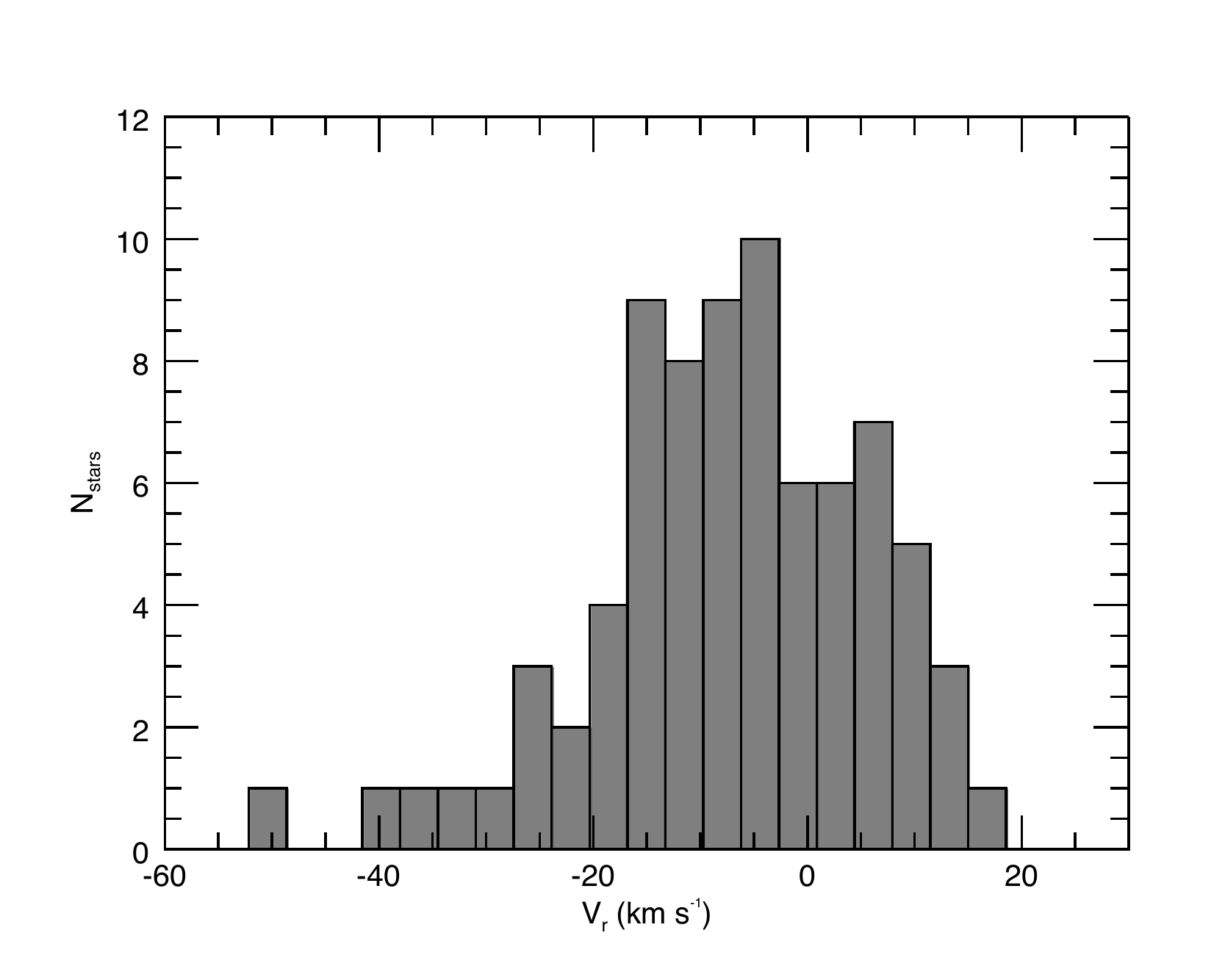}
\caption{Distribution of radial velocities of the observed stars. Stars that are known or candidate binaries are not included. }
\label{fig:Hist}
\end{figure}

As a large star forming region, we expect that the stars in Carina will have relatively similar radial velocities, but we find that there is a large $V_r$ dispersion across the nebula.  Figure \ref{fig:Hist} shows the distribution of radial velocities of the observed stars. Known and candidate binaries are not included. Overall, we find that $\langle V_{r} \rangle$ = -7.14 $\pm$ 13.10 km s$^{-1}$ for stars in Carina. The stars HD 93343 and HD 303304 have  $| V_r - \langle V_r \rangle | >$   30 km s$^{-1}$  and may be runaway stars from their respective birthplaces. Other stars might also be runaways if their proper motion is sufficiently high. \\

\begin{figure}
\centering
\includegraphics{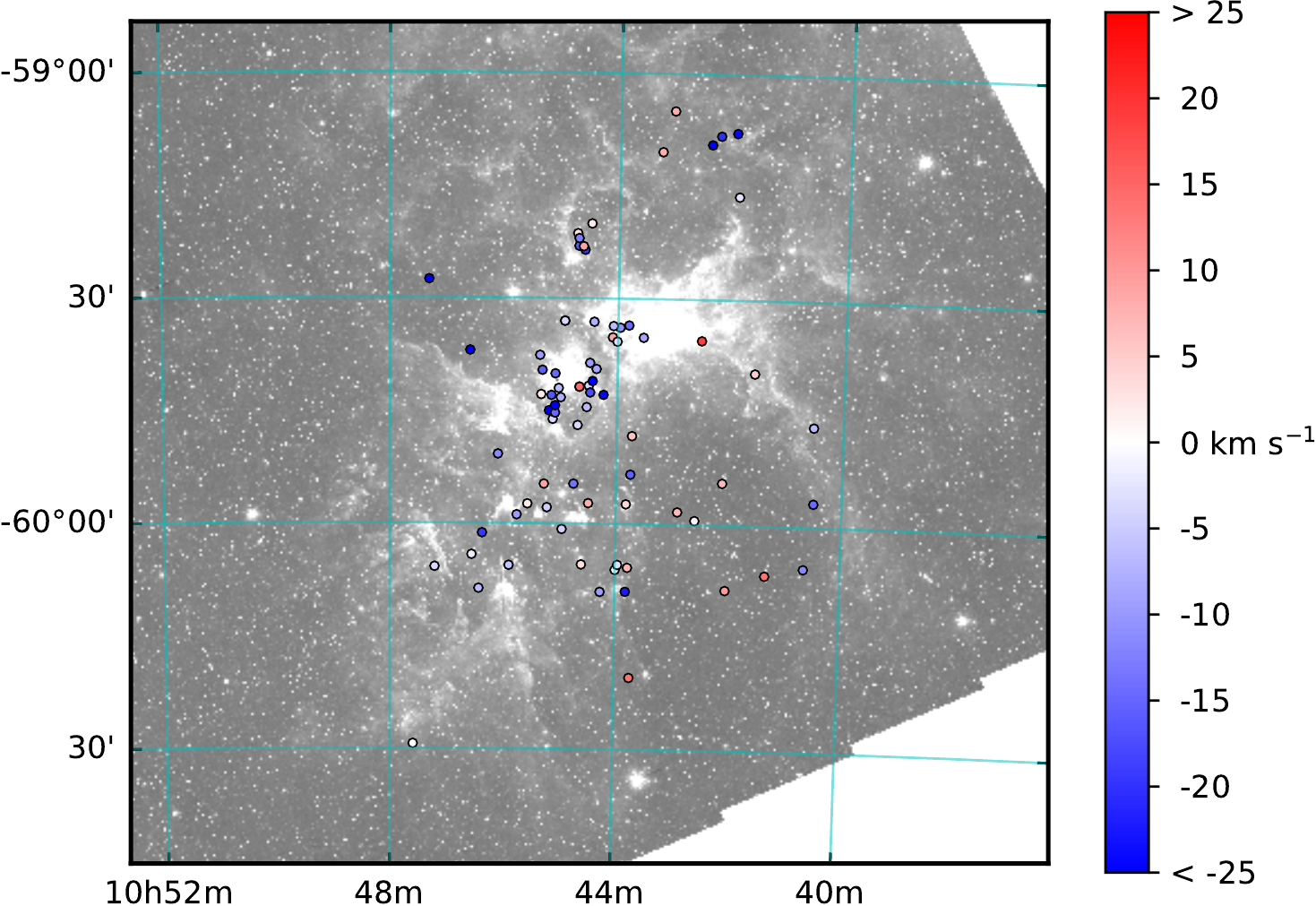}
\caption{Radial velocity distribution of unpaired stars throughout Carina. The color bar on the right corresponds to the radial velocities of the stars.}
\label{fig:VrP}
\end{figure}

We show the positional distribution of our radial velocity measurements in Figure \ref{fig:VrP}.  About half of the stars with high blueshifts are located far away from their main clusters, suggesting that they may also be runaways.  Similarly, the majority of the redshifted stars are located outside of the major clusters, with some being in low extinction ($A_V$) windows. These stars may also be a runaway population, a background OB population, or a mixture of both. \\

Carina lies in a complicated part of the Galaxy because it is near the tangent point of the Sagittarius-Carina spiral arm. The recombination and forbidden lines associated with the H II region occupy a larger range of radial velocities (-50 to 50 km s$^{-1}$) than does our OB population \citep{Damiani2016}. This isn't surprising, given that feedback from the OB stars drives expansion of the nebula. There is a persistent, two-peaked velocity structure in the nebular lines with ($V_{blue}$, $V_{red}$) = (-30,10) km s$^{-1}$ = $V_{cm}$ $\pm$ 20 km s$^{-1}$ \citep{Damiani2016}. The peak of the velocity distribution of our observed OB stars fits between these two peaks. The densest molecular clouds associated with the Carina Nebula are found toward the negative end of our OB velocities, near -20 km s$^{-1}$ \citep{Rebolledo2016}. Those authors conclude that higher velocity gas is associated with more distant regions of the Sagittarius-Carina arm of our Galaxy. \\

\begin{figure}
\centering
\includegraphics{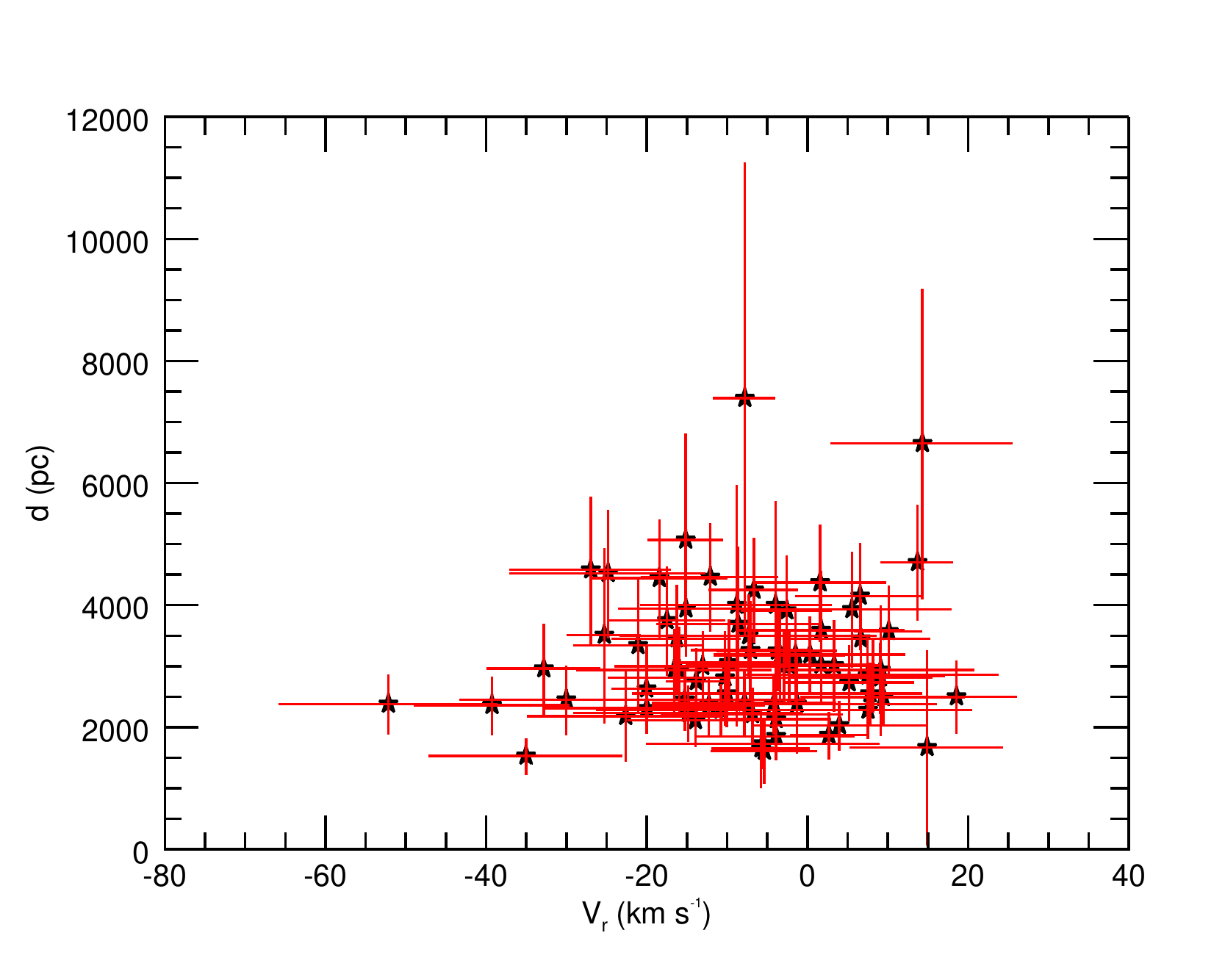}
\caption{Derived distances vs. radial velocities of the stars not in known or candidate binaries. }
\label{fig:VrD}
\end{figure}

Looking at the V$_r$ distribution in Figure \ref{fig:Hist}, we can imagine that the stars with V$_r$ $>$ 0 km s$^{-1}$ form a more distant component behind Carina, as suggested by \citet{Rebolledo2016}. Using the measured luminosities of the stars that are not in known or candidate binary systems, and the available data on the apparent magnitudes and $A_V$, we can estimate the distances to these stars. Our results can be found in Table \ref{tab:Photometry}. The bolometric corrections (BC) in column 4 are interpolated from \citet{Flower1996} and \citet{Torres2010}. The visible apparent magnitudes ($m_V$) in column 8 are from the SIMBAD database unless otherwise marked, with assumed uncertainty of 0.1, and the $A_V$ values in column 9 come from \citet{Povich2011a}. The distances in Table \ref{tab:Photometry} have large uncertainties, and we expect that Gaia parallaxes will resolve the distance uncertainties. In Figure \ref{fig:VrD}, we plot the derived distances against our measured V$_r$.  We find that there is no particular trend between the two quantities, suggesting that the stars with V$_r$ $>$ 0 km s$^{-1}$ may not be background stars. The stars LS 1763 and Tr 16-26 could be background stars, with d $>$ 6,000 pc; however the large uncertainties in their derived distances makes it difficult to be certain.\\

We find a large degree of $V_r$ dispersion, even in the OB populations associated with Tr 14, Tr 15, and Tr 16, so we estimated the dispersion we might expect for these massive clusters. We assumed the clusters are virialized and tested cluster masses of 1,000 M$_\odot$ or 10,000 M$_\odot$, and half-radii of 1 pc or 0.3 pc. This gives a range of V$_r$ dispersions from $\sim$2 km s$^{-1}$ to $\sim$10 km s$^{-1}$. The upper part of this range is consistent with our observed dispersion, and it is quite possible that the Carina clusters are not in virial equilibrium, which would increase the expected V$_r$ dispersion.  To check the status of virial equilibrium, we estimated the relaxation time of the individual clusters. We assumed that there are about 1,000 to 2,000 stars in the individual Carina clusters and that the stars have velocities within the clusters of around 1 to 10 km s$^{-1}$. We also assumed the clusters had radii ranging from 1 to 2 pc.  The relaxation time for these modeled clusters range from $\sim$1.4 to $\sim$52 Myr. We know that the clusters like Tr 14 and Tr 16 are about 10 Myr in age \citep{Getman2014}, so it is likely that these clusters are not relaxed, further bolstering the likelihood that they are not yet virialized.  \\

\section{Summary} \label{sec:sum}

Our goal with this paper was to create a catalog of spectroscopic parameters of a large sample of stars scattered throughout the Carina Nebula. Of the 128 B-type stars we spectroscopically classified in \citet{Alexander2016}, 82 of them had a S/N high enough to measure $T_{\rm eff}$, $\log g$, $V \sin i$, $V_{\rm r}$, $M_{\star}$, $\tau_{\star}$, $R_{\star}$, and $L_{\rm bol}$. With the recent and future public data releases from the Gaia-ESO Survey in mind, we also included the radial velocities of the B- and O-type stars in our sample, finding that about 29.2\% of our sample are SB1 candidates. We do not find any relationship between distance and $V_r$, implying that the high $V_r$ stars are probably not a collection of background stars viewed along the tangent of the Sagittarius-Carina arm of the Galaxy. Instead, we conclude that the Carina Nebula has not yet virialized.  \\

The authors would like to thank the referee for several helpful comments that improved this manuscript. The authors would like to thank Sara Martell, Angel Lopez-Sanchez, and Iraklis Konstantopoulos for their assistance with the AAT observations. M.V.M. is supported by the National Science Foundation under grant AST-1109247 and a Class of 1961 Professorship from Lehigh University. R.J.H. and M.V.M. have also received institutional support from Lehigh University. MSP is grateful to the NSF for support from awards AST-1411851 and CAREER-1454333. This research has made use of the WEBDA database, operated at the Department of Theoretical Physics and Astrophysics of the Masaryk University and the SIMBAD database, operated at CDS, Strasbourg, France.

\startlongtable
\begin{deluxetable}{lcccccccccc}
\tablecaption{Projected Rotational Velocities \label{tab:vsini}}
\tabletypesize{\scriptsize}
\tablehead{
\colhead{ID} & \colhead{S/N\textsubscript{13}}& \colhead{S/N\textsubscript{14}} & \colhead{$V \sin i_{4026}$} & \colhead{$\Delta V \sin i$}& \colhead{$V \sin i_{4388}$} & \colhead{$\Delta V \sin i$}& \colhead{$V \sin i_{4471}$} & \colhead{$\Delta V \sin i$} & \colhead{$V \sin i$} & \colhead{$\Delta V \sin i$}\\
\colhead{} & \colhead{} & \colhead{} & \colhead{($km~s^{-1}$)} & \colhead{($km~s^{-1}$)}& \colhead{($km~s^{-1}$)}& \colhead{($km~s^{-1}$)}& \colhead{($km~s^{-1}$)}& \colhead{($km~s^{-1}$)}& \colhead{($km~s^{-1}$)}&\colhead{($km~s^{-1}$)} 
}
\startdata
HD 93620  & 88      & 104     & 60      & 5       & 65      & 5       & 65      & 5       & 63  & 9  \\
HD 305606 & 63      & 122     & 55      & 5       & 45      & 5       & 50      & 5       & 50  & 9  \\
OBc89      & 62      & 113     & 35      & 5       & 35      & 5       & 35      & 5       & 35  & 9  \\
HD 93576  & 74      & 162     & 110     & 10      & 140     & 5       & 135     & 5       & 132 & 12 \\
HD 93501  & 89      & 135     & 185     & 7       & 225     & 7       & 220     & 7       & 210 & 12 \\
ERO 39      & 90      & 174     & 145     & 12      & 160     & 5       & 145     & 5       & 151 & 14 \\
OBc75      & 34      & 117     & 45      & 5       & 25      & 5       & 30      & 5       & 33  & 9  \\
Coll 228-81 & 56      & 127     & 90      & 5       & 90      & 5       & 100     & 5       & 93  & 9  \\ 
HD 305538 & 67      & 130     & 90      & 17      & 95      & 10      & 95      & 10      & 94  & 22 \\
HD 305528 & 82      & 130     & \nodata & \nodata & 70      & 7       & 65      & 5       & 67  & 9  \\
HD 305533 & 75      & 112     & 205     & 10      & 215     & 15      & 210     & 12      & 209 & 22 \\
LS 1866   & 51      & 138     & 60      & 10      & 60      & 5       & 60      & 5       & 60  & 12 \\
LS 1837   & 76      & 131     & 145     & 10      & 135     & 10      & 140     & 10      & 140 & 17 \\
HD 93097  & 107     & 123     & 185     & 7       & 190     & 7       & 200     & 7       & 192 & 12 \\
Coll 228-68 & 97      & 146     & 155     & 7       & 150     & 7       & 160     & 10      & 154 & 14 \\
Coll 228-48 & 80      & 126     & \nodata & \nodata & 170     & 5       & 175     & 5       & 173 & 7  \\
HD 93027  & 105     & 161     & 80      & 5       & 85      & 5       & 70      & 5       & 78  & 9  \\
Tr 16-20    & 95      & 115     & 55      & 7       & 80      & 10      & 50      & 5       & 58  & 13 \\
HD 305521 & 85      & 149     & 95      & 5       & 95      & 5       & 95      & 7       & 95  & 10 \\
HD 305452 & 79      & 172     & 35      & 5       & 35      & 5       & 40      & 5       & 37  & 9  \\
LS 1763   & 44      & 94      & 250     & 15      & 260     & 7       & 265     & 15      & 259 & 22 \\
HD 305535 & 133     & 134     & 175     & 10      & 200     & 5       & 200     & 5       & 195 & 12 \\
Coll 228-30 & 52      & 144     & 80      & 10      & 85      & 5       & 90      & 5       & 86  & 12 \\
LS 1813   & 76      & 145     & 95      & 10      & 80      & 5       & 80      & 10      & 84  & 15 \\
LS 1745   & 82      & 180     & 70      & 10      & 75      & 5       & 75      & 5       & 74  & 12 \\
LS 1760   & 50      & 144     & 35      & 5       & 45      & 5       & 45      & 5       & 42  & 9  \\
HD 92877  & 142     & 198     & 115     & 5       & 115     & 5       & 120     & 5       & 117 & 9  \\
Tr 16-16    & 72      & 123     & 100     & 12      & 130     & 10      & 135     & 10      & 123 & 19 \\
HD 305437 & 109     & 173     & 105     & 5       & 85      & 5       & 80      & 7       & 91  & 10 \\
HD 305443 & 48      & 145     & 45      & 7       & 30      & 2       & 30      & 2       & 32  & 8  \\
HD 305518 & 81      & 146     & 100     & 7       & 130     & 5       & 105     & 10      & 115 & 13 \\
HD 92644  & 106     & 179     & 185     & 7       & 190     & 7       & 195     & 10      & 189 & 14 \\
Tr 16-17    & 59      & 133     & 240     & 12      & 235     & 10      & 245     & 12      & 240 & 20 \\
HD 303225 & 85      & 162     & 75      & 10      & 80      & 7       & 85      & 7       & 81  & 14 \\
Tr 16-11    & 103     & 95      & \nodata & \nodata & 305     & 10      & 305     & 15      & 305 & 18 \\
HD 92937  & 108     & 191     & 115     & 7       & 145     & 5       & 130     & 7       & 132 & 11 \\
Tr 16-94    & 120     & 153     & 125     & 20      & 120     & 5       & 130     & 5       & 125 & 21 \\
Tr 14-30    & 50      & 138     & 50      & 5       & 55      & 5       & 50      & 5       & 52  & 9  \\
Tr 14-27    & 33      & 120     & 75      & 17      & 70      & 7       & 70      & 5       & 71  & 19 \\
HD 303189 & 69      & 131     & 95      & 10      & 100     & 5       & 100     & 7       & 99  & 13 \\
HD 303202 & 64      & 135     & 140     & 7       & 165     & 5       & 170     & 10      & 158 & 13 \\
HD 303297 & 71      & 151     & 60      & 5       & 60      & 5       & 50      & 5       & 57  & 9  \\
HD 92894  & 88      & 180     & 75      & 5       & 80      & 5       & 75      & 7       & 77  & 10 \\
HD 303296 & 86      & 139     & 75      & 5       & 80      & 5       & 65      & 5       & 73  & 9  \\
Tr 15-23    & 42      & 98      & 45      & 10      & 50      & 5       & 40      & 5       & 45  & 12 \\
HD 93026  & 86      & 150     & 95      & 5       & 105     & 5       & 95      & 7       & 99  & 10 \\
HD 93002  & 83      & 154     & 70      & 5       & 70      & 5       & 65      & 7       & 69  & 10 \\
LS 1822   & 73      & 107     & 60      & 12      & 75      & 7       & 65      & 5       & 67  & 15 \\
Tr 16-122   & 29      & 103     & \nodata & \nodata & 115     & 5       & 125     & 5       & 120 & 7  \\
Tr 14-19    & 28      & 77      & \nodata & \nodata & 240     & 10      & \nodata & \nodata & 240 & 10 \\
Tr 15-15    & \nodata & 117     & \nodata & \nodata & 140     & 10      & 130     & 10      & 135 & 14 \\
HD 93249 A  & \nodata & 130     & \nodata & \nodata & 110     & 5       & \nodata & \nodata & 110 & 5  \\
Tr 15-26    & 47      & 102     & 35      & 5       & 40      & 5       & 35      & 5       & 37  & 9  \\
Tr 16-31    & 70      & 106     & 225     & 12      & 225     & 7       & 210     & 7       & 219 & 16 \\
Tr 16-124   & 22      & 102     & \nodata & \nodata & 95      & 7       & 105     & 7       & 100 & 10 \\
Tr 16-4     & 37      & 88      & 70      & 17      & 100     & 5       & 100     & 5       & 96  & 18 \\
HD 303402 & 46      & 122     & \nodata & \nodata & 110     & 5       & 115     & 10      & 112 & 11 \\
Tr 16-2     & 48      & 99      & 250     & 17      & 220     & 12      & 245     & 7       & 239 & 20 \\
HD 93723  & 122     & 186     & 55      & 5       & 40      & 5       & 40      & 5       & 45  & 9  \\
Tr 16-3    & 70      & 134     & \nodata & \nodata & 65      & 5       & 55      & 5       & 60  & 7  \\
Tr 16-115   & 81      & 128     & 150     & 5       & 140     & 5       & 125     & 5       & 138 & 9  \\
OBc68      & 61      & 202     & 20      & 7       & 20      & 5       & 25      & 5       & 22  & 10 \\
OBc60      & 9       & 88      & \nodata & \nodata & 80      & 5       & 80      & 5       & 80  & 7  \\
OBc57      & 51      & 267     & 60      & 15      & 65      & 5       & 60      & 5       & 62  & 17 \\
OBc23      & 114     & 244     & 60      & 12      & 75      & 5       & 65      & 5       & 68  & 14 \\
Tr 16-246   & \nodata & 171     & \nodata & \nodata & 135     & 7       & 135     & 7       & 135 & 10 \\
Tr 14-18    & 30      & 166     & \nodata & \nodata & 125     & 5       & 145     & 5       & 135 & 7  \\
Tr 14-28    & 15      & 159     & 80      & 30      & 55      & 7       & \nodata & \nodata & 60  & 31 \\
Tr 14-22    & \nodata & 128     & \nodata & \nodata & 345     & 10      & \nodata & \nodata & 345 & 10 \\
Tr 15-19    & 18      & 134     & \nodata & \nodata & 215     & 7       & 235     & 5       & 227 & 9  \\
Tr 16-18    & 58      & 155     & \nodata & \nodata & 80      & 5       & 115     & 7       & 95  & 9  \\
Tr 14-29    & 31      & 186     & 180     & 12      & 125     & 10      & 165     & 15      & 154 & 22 \\
Tr 16-12    & 31      & 190     & \nodata & \nodata & 55      & 5       & 100     & 20      & 64  & 21 \\
Tr 15-21    & 12      & 126     & \nodata & \nodata & 290     & 15      & 290     & 7       & 290 & 17 \\
Tr 16-14    & 42      & 181     & \nodata & \nodata & 90      & 5       & \nodata & \nodata & 90  & 5  \\
Tr 15-9     & 33      & 118     & \nodata & \nodata & 180     & 5       & 170     & 5       & 175 & 7  \\
Tr 16-25    & \nodata & 156     & \nodata & \nodata & 45      & 5       & 55      & 5       & 50  & 7  \\
Tr 16-28    & 42      & 179     & 40      & 10      & 45      & 5       & 55      & 5       & 48  & 12 \\
Tr 16-55    & 4       & 152     & \nodata & \nodata & 115     & 5       & 145     & 5       & 130 & 7  \\
Tr 16-24    & 10      & 159     & \nodata & \nodata & 205     & 5       & 225     & 10      & 212 & 11 \\
Tr 16-74    & 47      & 160     & 105     & 7       & 100     & 5       & 110     & 5       & 105 & 10 \\
Tr 16-26   & 59      & \nodata & 195     & 37      & \nodata & \nodata & \nodata & \nodata & 195 & 37
\enddata
\end{deluxetable}

\startlongtable
\begin{deluxetable}{lcccccccccccc}
\tablecaption{Physical Parameters \label{tab:Teff}}
\tablehead{
\colhead{ID} & \colhead{$T_{\rm eff}$} & \colhead{$\Delta T_{\rm eff}$} & \colhead{$\log g$} & \colhead{$\Delta \log g$} & \colhead{$\tau$\textsubscript{$\star$}} & \colhead{$\Delta$$\tau$\textsubscript{$\star$}} & \colhead{M\textsubscript{$\star$}} & \colhead{$\Delta$M\textsubscript{$\star$} }& \colhead{R\textsubscript{$\star$}} &  \colhead{$\Delta$R\textsubscript{$\star$}} &  \colhead{L\textsubscript{bol}} & \colhead{$\Delta$L\textsubscript{bol}}\\
\colhead{} & \colhead{(K)} & \colhead{(K)} & \colhead{(dex)} &\colhead{(dex)} &\colhead{(Myr)} & \colhead{(Myr)} &\colhead{(M\textsubscript{$\odot$})} &\colhead{(M\textsubscript{$\odot$})}  &\colhead{(R\textsubscript{$\odot$})}  &\colhead{(R\textsubscript{$\odot$})} &  \colhead{(L\textsubscript{$\odot$})}& \colhead{(L\textsubscript{$\odot$})}
}
\startdata
HD 93620   & 19800 & 650  & 3.3  & 0.11 & 16.43 & 5.19  & 11.7  & 1.97  & 12.67 & 2.69 & 22160  & 9850   \\
HD 305606  & 21100 & 550  & 4.11 & 0.11 & 19.7  & 7.48  & 7.64  & 0.73  & 4.03  & 0.71 & 2890   & 1060   \\
OBc89      & 17700 & 650  & 3.26 & 0.12 & 22.8  & 8.13  & 10    & 1.8   & 12.27 & 2.82 & 13270  & 6410   \\
HD 93576   & 28700 & 350  & 3.72 & 0.11 & 7.65  & 0.85  & 18.07 & 2.46  & 9.71  & 1.9  & 57460  & 22700  \\
HD 93501   & 29700 & 350  & 3.87 & 0.1  & 7.26  & 0.3   & 17.15 & 1.9   & 7.96  & 1.36 & 44280  & 15330  \\
ERO 39      & 26700 & 2000 & 3.63 & 0.22 & 9.04  & 3.99  & 16.65 & 7.2   & 10.34 & 4.97 & 48780  & 49190  \\
OBc75      & 29700 & 350  & 3.89 & 0.13 & 7.25  & 0.45  & 16.87 & 2.25  & 7.72  & 1.68 & 41590  & 18250  \\
Coll 228-81 & 27000 & 1000 & 4.38 & 0.22 & 0     & 3.78  & 10.31 & 2.36  & 3.43  & 1.28 & 5620   & 4270   \\
HD 305538  & 25500 & 1350 & 4.27 & 0.16 & 2.52  & 6.1   & 10.13 & 1.87  & 3.86  & 1.07 & 5660   & 3370   \\
HD 305528  & 15000 & 300  & 3.3  & 0.11 & 40.66 & 12.37 & 7.21  & 0.96  & 9.95  & 1.94 & 4500   & 1790   \\
HD 305533  & 29400 & 650  & 4.11 & 0.14 & 5.62  & 3.14  & 14.24 & 1.72  & 5.5   & 1.23 & 20310  & 9290   \\
LS 1866    & 27200 & 1000 & 4.35 & 0.18 & 0     & 3.46  & 10.74 & 2.13  & 3.63  & 1.12 & 6460   & 4110   \\
LS 1837    & 26300 & 900  & 4.09 & 0.16 & 9.2   & 4.91  & 11.46 & 1.64  & 5.05  & 1.31 & 10960  & 5890   \\
HD 93097   & 26800 & 200  & 3.75 & 0.11 & 9.62  & 0.85  & 14.82 & 1.73  & 8.5   & 1.58 & 33440  & 12530  \\
Coll 228-68 & 27100 & 1000 & 4.15 & 0.2  & 6.51  & 5.49  & 11.8  & 2.11  & 4.78  & 1.56 & 11080  & 7400   \\
Coll 228-48 & 19000 & 1150 & 4.12 & 0.18 & 29.7  & 19.38 & 6.41  & 1.17  & 3.65  & 1.1  & 1560   & 1020   \\
HD 93027   & 29700 & 300  & 3.66 & 0.11 & 6.46  & 0.82  & 20.82 & 3.04  & 11.17 & 2.25 & 87180  & 35250  \\
Tr 16-20    & 18200 & 600  & 3.63 & 0.13 & 37.86 & 6.84  & 7.55  & 0.98  & 6.96  & 1.51 & 4780   & 2170   \\
HD 305521  & 29500 & 600  & 4.12 & 0.12 & 5.34  & 2.85  & 14.27 & 1.47  & 5.45  & 1.04 & 20170  & 7890   \\
HD 305452  & 20900 & 500  & 3.36 & 0.11 & 15.16 & 4.27  & 12.18 & 1.82  & 12.07 & 2.45 & 24950  & 10410  \\
LS 1763    & 22100 & 1250 & 3.42 & 0.18 & 14.21 & 6.56  & 12.8  & 3.71  & 11.55 & 4.14 & 28560  & 21490  \\
HD 305535  & 15000 & 500  & 3.15 & 0.11 & 31.56 & 9.28  & 8.36  & 1.41  & 12.74 & 2.7  & 7370   & 3280   \\
Coll 228-30 & 21200 & 750  & 3.82 & 0.13 & 23.18 & 3.81  & 8.95  & 1.36  & 6.09  & 1.39 & 6730   & 3220   \\
LS 1813    & 23200 & 800  & 3.84 & 0.13 & 17.46 & 2.81  & 10.52 & 1.62  & 6.46  & 1.47 & 10840  & 5170   \\
LS 1745    & 22800 & 550  & 3.57 & 0.11 & 16.27 & 2.59  & 11.74 & 1.37  & 9.30  & 1.74 & 21000  & 8130   \\
LS 1760    & 28600 & 400  & 3.63 & 0.11 & 7.13  & 1.03  & 19.44 & 3.01  & 11.17 & 2.30  & 74980  & 31150  \\
HD 92877   & 21800 & 650  & 3.65 & 0.11 & 18.97 & 3.24  & 10.76 & 1.2   & 8.12  & 1.48 & 13380  & 5140   \\
Tr 16-16    & 28000 & 1500 & 4.30  & 0.18 & 0.58  & 63.56 & 11.77 & 19.04 & 4.02  & 3.81 & 8920   & 17010  \\
HD 305437  & 29500 & 550  & 3.95 & 0.11 & 7.24  & 0.97  & 15.79 & 1.97  & 6.97  & 1.33 & 33000  & 12820  \\
HD 305443  & 22100 & 650  & 3.65 & 0.11 & 17.98 & 3.22  & 11.02 & 1.21  & 8.22  & 1.49 & 14480  & 5530   \\
HD 305518  & 26600 & 400  & 3.25 & 0.11 & 6.63  & 1.3   & 23.39 & 4.98  & 18.98 & 4.45 & 161970 & 76690  \\
HD 92644   & 29800 & 150  & 3.71 & 0.11 & 6.6   & 0.63  & 19.78 & 2.53  & 10.28 & 1.98 & 74800  & 28830  \\
Tr 16-17    & 28200 & 1700 & 3.82 & 0.2  & 8.48  & 1.85  & 15.78 & 5.15  & 8.09  & 3.25 & 37150  & 31200  \\
HD 303225  & 21300 & 1000 & 3.62 & 0.11 & 21.26 & 4.93  & 10.31 & 1.52  & 8.23  & 1.66 & 12530  & 5570   \\
Tr 16-11    & 23000 & 1500 & 3.74 & 0.27 & 16.83 & 5.34  & 11.09 & 3.23  & 7.44  & 3.49 & 13890  & 13560  \\
HD 92937   & 17700 & 700  & 3.20  & 0.11 & 20.44 & 6.85  & 10.63 & 1.94  & 13.56 & 2.97 & 16190  & 7550   \\
Tr 16-94    & 23200 & 600  & 4.10 & 0.11 & 14.28 & 5.5   & 8.98  & 0.88  & 4.42  & 0.78 & 5080   & 1880   \\
Tr 14-30    & 26100 & 1300 & 3.24 & 0.11 & 6.88  & 2.44  & 22.62 & 8.97  & 18.89 & 6.1  & 148600 & 100500 \\
Tr 14-27    & 28600 & 1000 & 4.60  & 0.13 & 0.00  & 0.00  & 9.30   & 1.76  & 2.53  & 0.62 & 3850   & 1970   \\
HD 303189  & 21400 & 800  & 3.45 & 0.11 & 16.44 & 5.25  & 11.69 & 2.00     & 10.66 & 2.28 & 21400  & 9700   \\
HD 303202  & 23100 & 850  & 3.52 & 0.12 & 13.94 & 3.21  & 12.91 & 1.99  & 10.33 & 2.24 & 27300  & 12520  \\
HD 303297  & 27800 & 2000 & 4.02 & 0.22 & 8.43  & 4.78  & 13.28 & 3.92  & 5.9   & 2.43 & 18640  & 16270  \\
HD 92894   & 25800 & 1250 & 3.55 & 0.16 & 9.41  & 2.69  & 16.62 & 4.00     & 11.33 & 3.48 & 51050  & 32950  \\
HD 303296  & 24300 & 750  & 3.67 & 0.13 & 12.79 & 2.96  & 13.01 & 2.24  & 8.73  & 2.08 & 23860  & 11740  \\
Tr 15-23    & 23500 & 750  & 4.07 & 0.13 & 14.71 & 5.53  & 9.36  & 1.15  & 4.67  & 1.00    & 5980   & 2660   \\
HD 93026   & 23700 & 650  & 3.9  & 0.11 & 16.41 & 1.8   & 10.52 & 1.25  & 6.02  & 1.13 & 10280  & 4010   \\
HD 93002   & 22900 & 750  & 3.65 & 0.11 & 15.42 & 3.3   & 11.71 & 1.6   & 8.47  & 1.66 & 17730  & 7330   \\
LS 1822    & 29800 & 150  & 4.07 & 0.11 & 5.96  & 1.74  & 14.86 & 1.09  & 5.89  & 0.97 & 24530  & 8120   \\
Tr 16-122   & 23200 & 1000 & 4.10  & 0.12 & 14.28 & 6.74  & 8.98  & 1.21  & 4.42  & 0.92 & 5080   & 2280   \\
Tr 14-19    & 15300 & 1000 & 3.5  & 0.14 & 58.5  & 26.4  & 6.31  & 1.41  & 7.39  & 2.04 & 2690   & 1640   \\
Tr 15-15    & 23000 & 850  & 3.45 & 0.11 & 13.27 & 3.88  & 13.41 & 2.48  & 11.42 & 2.52 & 32760  & 15240  \\
HD 93249 A   & 26800 & 250  & 3.26 & 0.1  & 6.59  & 1.35  & 23.51 & 5.93  & 18.82 & 4.55 & 163970 & 79660  \\
Tr 15-26    & 23700 & 900  & 3.75 & 0.14 & 14.96 & 3.56  & 11.59 & 2.10   & 7.51  & 1.91 & 15990  & 8500   \\
Tr 16-31    & 29600 & 450  & 3.89 & 0.11 & 7.33  & 0.5   & 16.73 & 2.11  & 7.68  & 1.47 & 40700  & 15730  \\
Tr 16-124   & 29500 & 600  & 4.02 & 0.22 & 6.73  & 3.02  & 14.94 & 2.84  & 6.25  & 2.24 & 26580  & 19200  \\
Tr 16-4     & 29700 & 400  & 4.13 & 0.16 & 4.98  & 3.38  & 14.4  & 1.66  & 5.41  & 1.32 & 20440  & 10080  \\
HD 303402  & 21700 & 850  & 3.29 & 0.12 & 12.3  & 3.78  & 14.05 & 2.98  & 14.05 & 3.46 & 39310  & 20330  \\
Tr 16-2     & 29500 & 650  & 4.1  & 0.16 & 5.73  & 3.29  & 14.39 & 1.97  & 5.60   & 1.43 & 21290  & 11080  \\
HD 93723   & 17800 & 300  & 3.73 & 0.11 & 38.62 & 4.07  & 7.12  & 0.5   & 6.03  & 0.98 & 3270   & 1080   \\
Tr 16-3    & 29700 & 350  & 3.78 & 0.11 & 7.05  & 0.63  & 18.5  & 2.45  & 9.17  & 1.78 & 58770  & 23000  \\
Tr 16-115   & 29800 & 250  & 3.57 & 0.11 & 5.9   & 0.77  & 23.37 & 4.05  & 13.13 & 2.82 & 122020 & 52620  \\
OBc68      & 29100 & 1000 & 4.37 & 0.16 & 0.00     & 1.81  & 11.89 & 2.36  & 3.73  & 1.07 & 8950   & 5270   \\
OBc60      & 25100 & 950  & 3.86 & 0.16 & 12.68 & 2.13  & 11.9  & 2.22  & 6.71  & 1.89 & 16030  & 9350   \\
OBc57      & 21800 & 850  & 4.17 & 0.14 & 14.2  & 10.73 & 7.9   & 0.99  & 3.82  & 0.86 & 2970   & 1420   \\
OBc23      & 16600 & 700  & 3.44 & 0.13 & 38.12 & 15.65 & 7.52  & 1.35  & 8.65  & 2.09 & 5100   & 2610   \\
Tr 16-246   & 28100 & 2000 & 3.87 & 0.18 & 8.58  & 2.12  & 14.9  & 4.69  & 7.42  & 2.75 & 30820  & 24520  \\
Tr 14-18    & 22600 & 650  & 4.1  & 0.11 & 15.91 & 6.12  & 8.62  & 0.86  & 4.33  & 0.77 & 4390   & 1640   \\
Tr 14-28    & 28900 & 1000 & 4.24 & 0.16 & 2.19  & 3.73  & 13.05 & 2.05  & 4.54  & 1.2  & 12890  & 7040   \\
Tr 14-22    & 20800 & 1350 & 3.53 & 0.16 & 21.28 & 7.89  & 10.4  & 2.5   & 9.17  & 2.83 & 14130  & 9480   \\
Tr 15-19    & 16000 & 750  & 3.83 & 0.14 & 62.94 & 12.75 & 5.74  & 0.87  & 4.82  & 1.15 & 1370   & 700    \\
Tr 16-18    & 29500 & 650  & 4.26 & 0.13 & 1.35  & 2.81  & 13.5  & 1.65  & 4.51  & 0.95 & 13810  & 5950   \\
Tr 14-29    & 23100 & 1050 & 4.14 & 0.14 & 12.84 & 8.83  & 8.79  & 1.27  & 4.18  & 0.99 & 4460   & 2260   \\
Tr 16-12    & 29500 & 650  & 4.12 & 0.14 & 5.34  & 3.22  & 14.27 & 1.71  & 5.45  & 1.22 & 20170  & 9200   \\
Tr 15-21    & 15200 & 200  & 3.92 & 0.11 & 74.13 & 4.75  & 4.96  & 0.44  & 4.04  & 0.7  & 780    & 270    \\
Tr 16-14    & 27000 & 2000 & 4.11 & 0.18 & 7.68  & 6.32  & 11.89 & 2.87  & 5.03  & 1.68 & 12060  & 8810   \\
Tr 15-9     & 17400 & 500  & 3.85 & 0.11 & 47.55 & 7.84  & 6.44  & 0.68  & 4.99  & 0.9  & 2050   & 780    \\
Tr 16-25    & 20100 & 500   & 4.05 & 0.14 & 25.67 & 5.63  & 7.2   & 0.54  & 4.19  & 0.84 & 2580   & 1030   \\
Tr 16-28    & 29200 & 400  & 4.36 & 0.12 & 0.00     & 1.11  & 12.07 & 1.54  & 3.8   & 0.77 & 9420   & 3870   \\
Tr 16-55    & 25000 & 1100 & 4.15 & 0.14 & 9.28  & 6.93  & 10.22 & 1.45  & 4.45  & 1.04 & 6950   & 3480   \\
Tr 16-24    & 22300 & 800  & 4.1  & 0.13 & 16.75 & 7.62  & 8.43  & 1.03  & 4.28  & 0.91 & 4070   & 1830   \\
Tr 16-74    & 29300 & 850  & 4.22 & 0.13 & 2.72  & 3.40   & 13.55 & 1.74  & 4.73  & 1.01 & 14800  & 6580   \\
Tr 16-26   & 24800 & 2000 & 3.44 & 0.22 & 10.19 & 4.98  & 16.11 & 7.19  & 12.66 & 6.18 & 54430  & 56050
\enddata
\end{deluxetable}

\startlongtable
\begin{deluxetable}{lccccl}
\tablecaption{Radial Velocity Measurements of B-type stars \label{tab:Vr}}
\tablehead{\colhead{ID} & \colhead{$V_{r,2013}$} & \colhead{$\Delta V_{r,2013}$}& \colhead{$V_{r,2014}$} & \colhead{$\Delta V_{r,2014}$} & \colhead{Notes} \\
\colhead{} & \colhead{km s$^{-1}$} & \colhead{km s$^{-1}$} & \colhead{km s$^{-1}$} & \colhead{km s$^{-1}$} & \colhead{}
}
\startdata
HD 93620  & 21.18   & 4.69    & -38.98  & 1.96    & SB1c    \\
HD 305606 & 12.99   & 5.43    & -16.53  & 3.58    & SB1c    \\
OBc89      & 6.68    & 1.51    & -11.30  & 2.42    & SB1c    \\
HD 93576  & 7.90    & 6.38    & -73.85  & 13.35   & SB1c , SB\tablenotemark{b}\\
HD 93501  & -23.51  & 9.09    & -16.49  & 8.50    & \nodata \\
ERO 39      & 1.00    & 6.73    & 53.81   & 9.70    & SB1c    \\
OBc75      & 8.65    & 5.59    & -8.80   & 3.57    & \nodata \\
Coll 228-81 & 2.91    & 4.42    & -14.46  & 4.16    & \nodata \\   
HD 305538 & 23.84   & 6.13    & -47.87  & 8.99    & SB1c , Double/Multiple Star\tablenotemark{b}\\
HD 305528 & 19.61   & 6.18    & -1.44   & 1.94    & \nodata \\
HD 305533 & 4.91    & 2.89    & -15.88  & 14.18   & \nodata \\
LS 1866   & 4.15    & 1.36    & -14.88  & 6.41    & \nodata \\
LS 1837   & 0.30    & 9.13    & -20.83  & 11.19   & \nodata \\
HD 93097  & 17.07   & 7.79    & -3.77   & 3.80    & \nodata \\
Coll 228-68 & 9.32    & 17.23   & -11.91  & 2.59    & \nodata \\
Coll 228-48 & -13.76  & 5.62    & -31.50  & 10.73   & \nodata \\
HD 93027  & 10.26   & 2.16    & -12.01  & 4.13    & SB1c    \\
Tr 16-20    & 10.83   & 4.82    & -21.31  & 3.58    & SB1c    \\
HD 305521 & 12.41   & 6.38    & -4.48   & 8.74    & \nodata \\
HD 305452 & 7.83    & 2.69    & 12.41   & 3.22    & \nodata \\
LS 1763   & 15.61   & 10.17   & 12.99   & 4.82    & \nodata \\
HD 305535 & 13.00   & 6.46    & 1.95    & 7.10    & \nodata \\
Coll 228-30 & 11.26   & 6.20    & -14.23  & 4.51    & \nodata \\
LS 1813   & -9.73   & 3.38    & -22.18  & 5.66    & \nodata \\
LS 1745   & 1.38    & 8.06    & -25.55  & 2.64    & \nodata \\
LS 1760   & 12.58   & 6.82    & 0.53    & 4.00    & \nodata \\
HD 92877  & -2.57   & 5.28    & -25.32  & 6.38    & Double/Multiple Star\tablenotemark{b}\\
Tr 16-16    & 4.13    & 3.45    & 25.66   & 8.81    & \nodata \\
HD 305437 & 44.78   & 11.41   & -6.23   & 5.45    & SB1c    \\
HD 305443 & -3.42   & 4.26    & -26.84  & 7.17    & \nodata \\
HD 305518 & 5.19    & 6.36    & 5.87    & 10.61   & \nodata \\
HD 92644  & 7.57    & 6.89    & -22.20  & 14.32   & \nodata \\
Tr 16-17    & 3.32    & 4.95    & -11.27  & 1.42    & \nodata \\
HD 303225 & 16.34   & 7.04    & -5.96   & 3.91    & \nodata \\
Tr 16-11    & -11.39  & 9.05    & -6.19   & 7.65    & \nodata \\
HD 92937  & 32.95   & 4.36    & 4.15    & 6.13    & \nodata \\
Tr 16-94    & -28.43  & 10.05   & -41.64  & 6.56    & \nodata \\
Tr 14-30    & -5.24   & 9.86    & -12.04  & 1.73    & \nodata \\
Tr 14-27    & 8.20    & 5.10    & -19.34  & 3.27    & SB1c    \\
HD 303189 & -16.96  & 10.16   & -32.66  & 6.65    & \nodata \\
HD 303202 & \nodata & \nodata & -2.57   & 5.61    & \nodata \\
HD 303297 & -4.68   & 3.83    & -60.11  & 1.22    & SB1c    \\
HD 92894  & -14.83  & 7.55    & -27.35  & 2.47    & \nodata \\
HD 303296 & -34.34  & 6.69    & -31.31  & 2.11    & \nodata \\
Tr 15-23    & 11.47   & 2.95    & -7.86   & 2.85    & SB1c    \\
HD 93026  & 17.47   & 12.31   & -2.40   & 4.06    & \nodata \\
HD 93002  & 21.83   & 10.39   & -5.48   & 11.70   & \nodata \\
LS 1822   & 21.93   & 13.59   & -46.34  & 3.33    & SB1c    \\
Tr 16-122   & 3.91    & 8.40    & -19.56  & 7.67    & \nodata \\
Tr 14-19    & 10.24   & 0.66    & -3.46   & 0.85    & SB1c    \\
Tr 15-15    & \nodata & \nodata & -17.49  & 7.25    & \nodata \\
HD 93249 A  & \nodata & \nodata & 3.32    & 1.77    & \nodata \\
Tr 15-26    & 12.54   & 9.91    & -9.12   & 3.03    & \nodata \\
Tr 16-31    & 11.62   & 11.72   & -11.01  & 1.63    & \nodata \\
Tr 16-124   & 18.23   & 9.40    & 0.03    & 6.81    & \nodata \\
Tr 16-4     & 3.33    & 5.53    & -17.55  & 4.72    & \nodata \\
HD 303402 & -13.40  & 6.87    & -40.56  & 7.25    & \nodata \\
Tr 16-2     & 15.10   & 2.82    & -6.62   & 1.00    & SB1c    \\
HD 93723  & 14.55   & 4.29    & -17.17  & 3.71    & SB1c    \\
Tr 16-3    & 5.20    & 5.64    & -12.67  & 4.86    & \nodata \\
Tr 16-115   & 16.13   & 5.03    & -13.00  & 6.58    & \nodata \\
OBc68      & -2.90   & 7.15    & -36.22  & 2.01    & SB1c    \\
OBc60      & 2.45    & 4.12    & -22.84  & 5.71    & \nodata \\
OBc57      & -5.48   & 1.24    & -22.89  & 1.24    & SB1c    \\
OBc23      & 11.61   & 2.88    & -13.10  & 2.32    & SB1c    \\
Tr 16-246   & \nodata & \nodata & -25.29  & 4.50    & \nodata \\
Tr 14-18    & 0.21    & 15.14   & -21.76  & 2.86    & \nodata \\
Tr 14-28    & 25.73   & 7.25    & -18.06  & 2.05    & SB1c    \\
Tr 14-22    & \nodata & \nodata & -15.16  & 4.64    & \nodata \\
Tr 15-19    & -22.81  & 3.74    & -17.26  & 1.96    & \nodata \\
Tr 16-18    & -9.51   & 4.28    & -27.37  & 7.36    & \nodata \\
Tr 14-29    & -7.69   & 5.43    & -6.02   & 4.94    & ERO 21\tablenotemark{c} \\
Tr 16-12    & -1.22   & 3.64    & -18.35  & 4.99    & \nodata \\
Tr 15-21    & 27.71   & 5.41    & -8.84   & 8.59    & \nodata \\
Tr 16-14    & 28.26   & 6.47    & -8.85   & 1.16    & SB1c    \\
Tr 15-9     & -6.58   & 8.59    & -19.51  & 6.56    & \nodata \\
Tr 16-25    & \nodata & \nodata & -30.55  & 4.81    & SB2c\tablenotemark{a}\\
Tr 16-28    & 3.84    & 11.23   & -23.92  & 2.94    & \nodata \\
Tr 16-55    & -11.13  & 6.85    & -21.44  & 4.10    & \nodata \\
Tr 16-24    & -28.27  & 9.73    & -31.76  & 8.87    & \nodata \\
Tr 16-74    & -8.92   & 6.94    & -24.24  & 9.91    & \nodata \\
Tr 16-26   & -7.80   & 3.79    & \nodata & \nodata & \nodata
\enddata
\tablenotetext{a}{See Section \ref{sec:spp} for details} 
\tablenotetext{b}{Simbad}
\tablenotetext{c}{\citealt{Sexton2015}}
\end{deluxetable}

\startlongtable
\begin{deluxetable}{lccccl}
\tablecaption{Radial Velocity Measurements of O-type stars \label{tab:Vr_O}}
\tablehead{\colhead{ID} & \colhead{$V_{r,2013}$} & \colhead{$\Delta V_{r,2013}$}& \colhead{$V_{r,2014}$} & \colhead{$\Delta V_{r,2014}$} & \colhead{Notes} \\
\colhead{} & \colhead{km s$^{-1}$} & \colhead{km s$^{-1}$} & \colhead{km s$^{-1}$} & \colhead{km s$^{-1}$} & \colhead{}
}
\startdata
Tr 16-127   & 5.39    & 6.44    & -12.22  & 4.93    & \nodata \\
HD 305438 & 4.23    & 2.85    & -12.95  & 1.67  & SB1c    \\
HD 303316 & 16.58   & 2.66    & -7.67   & 0.89  & SB1c    \\
HD 93028  & 31.53   & 2.52    & 23.83   & 2.06  & SB\tablenotemark{a} \\
HD 303312 & \nodata & \nodata & -0.24   & 0.81  & EB\tablenotemark{a} \\
HD 305556 & 22.48   & 3.77    & 4.93    & 2.34  & \nodata \\
Tr 14-21    & \nodata & \nodata & -6.09   & 5.50  & SB\tablenotemark{b} \\
HD 93128   & 24.56   & 6.18    & -11.44  & 4.09  & SB1c, SB\tablenotemark{c}   \\
LS 1821   & 11.56   & 8.54    & -17.40  & 8.51  & \nodata \\
HD 93130  & -71.22  & 8.26    & -50.85  & 1.96  & EB\tablenotemark{a}\\
HD 305536 & 20.32   & 4.97    & 13.20   & 3.01  & SB\tablenotemark{a} \\
HD 305523 & 19.51   & 5.85    & -3.96   & 2.92  & \nodata \\
HD 93204  & -1.71   & 8.45    & -12.02  & 7.50  & \nodata \\
HD 93222  & 12.10   & 4.10    & -6.79   & 2.26  & \nodata \\
HD 303311 & 29.86   & 1.14    & -15.21  & 8.17  & SB1c, Double/Multiple Star\tablenotemark{a}    \\
Tr 16-100   & -8.70   & 8.43    & 0.87    & 4.92  & \nodata \\
HD 305524 & -14.16  & 3.34    & -13.52  & 1.32  & \nodata \\
LS 1865   & 15.13   & 4.70    & -14.35  & 3.76  & SB1c    \\
Tr 16-23    & -10.80  & 9.95    & -18.90  & 10.11 & \nodata \\
HD 303308 & -11.90  & 14.55   & -18.65  & 3.55  & \nodata \\
Tr 16-22    & 5.87    & 5.47    & -13.66  & 4.04  & \nodata \\
HD 93343  & -45.61  & 9.27    & -58.72  & 9.90  & \nodata \\
HD 305532 & 8.43    & 5.88    & -5.04   & 3.34  & \nodata \\
FO 15     & -0.37   & 3.68    & -102.87 & 6.22  & SB1c, EB\tablenotemark{a}    \\
HD 305525 & -9.44   & 5.24    & -15.11  & 4.56  & \nodata \\
LS 1892   & 0.58    & 1.12    & -18.91  & 1.32  & SB1c    \\
LS 1893   & 1.43    & 4.70    & -14.76  & 2.83  & \nodata \\
HD 305539 & 8.32    & 6.96    & -12.88  & 2.97  & \nodata \\
HD 303304 & -40.47  & 1.71    & -38.07  & 9.45  & \nodata \\
HD 93632  & 2.79    & 7.24    & -11.20  & 4.36  & \nodata \\
LS 1914   & 13.91   & 6.39    & -9.07   & 0.37  & SB1c    \\
HD 305619 & 5.87    & 4.04    & -16.81  & 3.17  & SB1c    \\
HD 305599 & 28.09   & 7.61    & -10.35  & 2.96  & SB1c 
\enddata
\tablecomments{EB - eclipsing binary; SB - spectroscopic binary}
\tablenotetext{a}{Simbad}
\tablenotetext{b}{WEBDA}
\tablenotetext{c}{\citealt{Levato1991}}
\end{deluxetable}

\startlongtable
\begin{deluxetable}{lccccccccc}
\tablecaption{Radial Velocity Measurements of SB2s \label{tab:SB2}}
\tablehead{\colhead{ID} & \colhead{$V_{r,p,2013}$} & \colhead{$\Delta V_{r,p,2013}$}& \colhead{$V_{r,p,2014}$} & \colhead{$\Delta V_{r,p,2014}$} & \colhead{$V_{r,s,2013}$} & \colhead{$\Delta V_{r,s,2013}$}& \colhead{$V_{r,s,2014}$} & \colhead{$\Delta V_{r,s,2014}$} \\
\colhead{} & \colhead{km s$^{-1}$} & \colhead{km s$^{-1}$} & \colhead{km s$^{-1}$} & \colhead{km s$^{-1}$} & \colhead{km s$^{-1}$} & \colhead{km s$^{-1}$}& \colhead{km s$^{-1}$} & \colhead{km s$^{-1}$}
}
\startdata
HD 303313 & -120.32 & 9.94  & -69.32  & 8.88  & 125.76 & 13.22 & 34.82  & 4.04 \\
HD 93056  & 14.08   & 4.27  & -96.40  & 10.81 & \nodata  & \nodata  & 85.60  & 8.56 \\
HD 305522 & 23.28   & 9.32  & 52.60   & 5.23  & -31.70 & 13.86 & -67.47 & 3.17 \\
LS 1840   & -16.51  & 7.29  & -14.19  & 8.83  & 44.88  & 12.22 & \nodata & \nodata \\
HD 305534 & -142.93 & 10.71 & -117.25 & 19.91 & 145.33 & 33.07 & 96.26  & 31.15 \\
Tr 16-9   & 22.09   & 2.24  & -77.82  & 7.78  & -31.48 & 6.44  & 66.80  & 2.52 \\
Tr 16-1   & -44.65  & 3.08  & -129.93 & 12.48 & 37.23  & 0.67  & 86.52  & 22.96 \\
OBc49     & -48.48  & 11.55 & -83.57  & 9.72  & 86.70  & 22.22 & 45.22  & 12.78 \\
HD 92607 & -185.04 & 10.79 & -38.16  & 9.23    & 206.76 & 23.44 & 20.98   & 8.91 \\
Tr 16-10 & 29.77   & 7.21  & \nodata & \nodata & -14.26 & 8.35  & \nodata & \nodata  \\
Tr 16-21 & 44.01   & 10.91 & 99.07   & 3.99    & -58.98 & 10.60 & -87.22  & 2.23        
\enddata
\end{deluxetable}

\startlongtable
\begin{deluxetable}{lcccccccccc}
\tablecaption{Photometry and Derived Distances \label{tab:Photometry}}
\tablehead{\colhead{ID} & \colhead{$M_{bol}$} & \colhead{$\Delta M_{bol}$}& \colhead{$BC$} & \colhead{$\Delta BC$} & \colhead{$M_V$} & \colhead{$\Delta M_V$}& \colhead{$V$} & \colhead{$A_V$} & \colhead{$d~(pc)$} & \colhead{$\Delta d~(pc)$}
}
\colnumbers
\startdata
Coll 228-30 & -4.84   & 1.20    & -2.02   & 0.18    & -2.82                  & 1.21    & 10.8\tablenotemark{a}  & 1.10    & 3190    & 770     \\
Coll 228-48 & -3.25   & 1.63    & -1.78   & 0.31    & -1.47                  & 1.66    & 11.00\tablenotemark{a} & 0.78    & 2180    & 730     \\
Coll 228-68 & -5.38   & 1.67    & -2.61   & 0.22    & -2.77                  & 1.68    & 10.16\tablenotemark{a} & 1.05    & 2380    & 800     \\
Coll 228-81 & -4.64   & 1.90    & -2.60   & 0.22    & -2.04                  & 1.91    & 10.89\tablenotemark{a} & 1.85    & 1650    & 630     \\
HD 303189   & -6.10   & 1.13    & -2.04   & 0.19    & -4.05                  & 1.15    & 10.10                  & 0.88    & 4520    & 1040    \\
HD 303202   & -6.36   & 1.15    & -2.22   & 0.20    & -4.14                  & 1.16    & 9.80                   & 0.98    & 3910    & 910     \\
HD 303225   & -5.51   & 1.11    & -2.03   & 0.24    & -3.48                  & 1.14    & 9.74                   & 1.04    & 2730    & 620     \\
HD 303296   & -6.21   & 1.23    & -2.34   & 0.17    & -3.88                  & 1.24    & 9.50                   & 1.02    & 2960    & 740     \\
HD 303304   & \nodata & \nodata & \nodata & \nodata & -5.00\tablenotemark{b} & 0.30\tablenotemark{b}  & 9.71                   & 2.85    & 2360    & 470     \\
HD 303308   & \nodata & \nodata & \nodata & \nodata & -5.50\tablenotemark{b} & 0.30\tablenotemark{b}  & 8.17                   & 1.72    & 2450    & 490     \\
HD 303402   & -6.76   & 1.29    & -2.07   & 0.20    & -4.68                  & 1.31    & 10.69                  & 2.07    & 4580    & 1200    \\
HD 305443   & -5.67   & 0.95    & -2.12   & 0.15    & -3.56                  & 0.97    & 10.60                  & 1.18    & 3940    & 770     \\
HD 305452   & -6.26   & 1.04    & -1.99   & 0.12    & -4.27                  & 1.05    & 9.56                   & 1.07    & 3570    & 750     \\
HD 305518   & -8.29   & 1.18    & -2.56   & 0.09    & -5.73                  & 1.19    & 9.72                   & 2.48    & 3930    & 940     \\
HD 305521   & -6.03   & 0.98    & -2.84   & 0.13    & -3.19                  & 0.99    & 9.81                   & 1.46    & 2030    & 400     \\
HD 305523   & \nodata & \nodata & \nodata & \nodata & -5.50\tablenotemark{b} & 0.30\tablenotemark{b}  & 8.50                   & 1.97    & 2550    & 510     \\
HD 305524   & \nodata & \nodata & \nodata & \nodata & -5.00\tablenotemark{b} & 0.30\tablenotemark{b}  & 9.32                   & 2.12    & 2750    & 550     \\
HD 305525   & \nodata & \nodata & \nodata & \nodata & -5.50\tablenotemark{b} & 0.30\tablenotemark{b}  & 10.00                  & 3.64    & 2360    & 470     \\
HD 305528   & -4.40   & 0.99    & -1.24   & 0.11    & -3.16                  & 1.00    & 10.32                  & 1.24    & 2810    & 570     \\
HD 305532   & \nodata & \nodata & \nodata & \nodata & -5.00\tablenotemark{b} & 0.30\tablenotemark{b}  & 10.20                  & 2.80    & 3020    & 610     \\
HD 305533   & -6.04   & 1.14    & -2.83   & 0.14    & -3.21                  & 1.15    & 10.32                  & 2.33    & 1730    & 400     \\
HD 305535   & -4.94   & 1.11    & -1.24   & 0.18    & -3.70                  & 1.13    & 9.39                   & 0.82    & 2840    & 640     \\
HD 305539   & \nodata & \nodata & \nodata & \nodata & -4.65\tablenotemark{b} & 0.30\tablenotemark{b}  & 9.90                   & 2.16    & 3010    & 600     \\
HD 305556   & \nodata & \nodata & \nodata & \nodata & -6.00\tablenotemark{b} & 0.30\tablenotemark{b}  & 8.95                   & 1.59    & 4700    & 940     \\
HD 92644    & -7.45   & 0.96    & -2.87   & 0.03    & -4.58                  & 0.96    & 8.88                   & 0.74    & 3500    & 680     \\
HD 92877    & -5.59   & 0.96    & -2.08   & 0.16    & -3.50                  & 0.97    & 8.50                   & 0.38    & 2110    & 410     \\
HD 92894    & -7.04   & 1.61    & -2.48   & 0.28    & -4.56                  & 1.64    & 9.53                   & 1.47    & 3340    & 1090    \\
HD 92937    & -5.79   & 1.17    & -1.62   & 0.20    & -4.17                  & 1.18    & 8.95                   & 1.13    & 2500    & 590     \\
HD 93002    & -5.89   & 1.03    & -2.20   & 0.17    & -3.69                  & 1.05    & 9.71                   & 1.12    & 2860    & 600     \\
HD 93026    & -5.30   & 0.98    & -2.28   & 0.15    & -3.02                  & 0.99    & 9.67                   & 0.90    & 2280    & 450     \\
HD 93097    & -6.58   & 0.94    & -2.58   & 0.04    & -4.00                  & 0.94    & 9.76                   & 1.07    & 3450    & 650     \\
HD 93204    & \nodata & \nodata & \nodata & \nodata & -5.00\tablenotemark{b} & 0.30\tablenotemark{b}  & 8.42                   & 1.70    & 2210    & 440     \\
HD 93222    & \nodata & \nodata & \nodata & \nodata & -5.00\tablenotemark{b} & 0.30\tablenotemark{b}  & 8.10                   & 1.74    & 1870    & 380     \\
HD 93249    & -8.31   & 1.21    & -2.58   & 0.06    & -5.73                  & 1.22    & 8.20                   & 1.53    & 3010    & 740     \\
HD 93343    & \nodata & \nodata & \nodata & \nodata & -4.65\tablenotemark{b} & 0.30\tablenotemark{b}  & 9.56                   & 2.33    & 2380    & 480     \\
HD 93501    & -6.89   & 0.87    & -2.86   & 0.08    & -4.02                  & 0.87    & 9.09                   & 1.30    & 2310    & 400     \\
HD 93632    & \nodata & \nodata & \nodata & \nodata & -5.50\tablenotemark{b} & 0.30\tablenotemark{b}  & 9.10                   & 2.72    & 2380    & 480     \\
LS 1745     & -6.08   & 0.97    & -2.19   & 0.13    & -3.89                  & 0.98    & 9.92                   & 0.56    & 4460    & 880     \\
LS 1760     & -7.46   & 1.04    & -2.76   & 0.09    & -4.70                  & 1.04    & 10.61                  & 2.22    & 4150    & 870     \\
LS 1763     & -6.41   & 1.88    & -2.12   & 0.30    & -4.29                  & 1.90    & 11.18                  & 1.36    & 6650    & 2540    \\
LS 1813     & -5.36   & 1.19    & -2.23   & 0.19    & -3.13                  & 1.21    & 10.43                  & 1.22    & 2940    & 710     \\
LS 1821     & \nodata & \nodata & \nodata & \nodata & -4.30\tablenotemark{b} & 0.30\tablenotemark{b}  & 9.31                   & 1.22    & 3010    & 600     \\
LS 1837     & -5.37   & 1.34    & -2.53   & 0.20    & -2.84                  & 1.36    & 10.52                  & 1.11    & 2810    & 770     \\
LS 1866     & -4.80   & 1.59    & -2.62   & 0.22    & -2.18                  & 1.61    & 10.81                  & 1.95    & 1610    & 520     \\
LS 1893     & \nodata & \nodata & \nodata & \nodata & -4.00\tablenotemark{b} & 0.30\tablenotemark{b}  & 10.80                  & 1.66    & 4250    & 850     \\
OBc60       & -5.78   & 1.46    & -2.42   & 0.21    & -3.37                  & 1.47    & \nodata                & 4.90    & \nodata & \nodata \\
OBc75       & -6.82   & 1.10    & -2.86   & 0.08    & -3.96                  & 1.10    & \nodata                & 2.50    & \nodata & \nodata \\
Tr 14-18    & -4.38   & 0.93    & -2.17   & 0.15    & -2.21                  & 0.95    & 11.90                  & 2.32    & 2280    & 430     \\
Tr 14-22    & -5.65   & 1.68    & -1.98   & 0.33    & -3.67                  & 1.71    & 12.23                  & 2.37    & 5070    & 1740    \\
Tr 14-29    & -4.39   & 1.27    & -2.22   & 0.24    & -2.18                  & 1.29    & 11.94                  & \nodata & \nodata & \nodata \\
Tr 14-30    & -8.20   & 1.69    & -2.51   & 0.29    & -5.69                  & 1.72    & 10.07                  & 2.92    & 3690    & 1270    \\
Tr 15-15    & -6.56   & 1.16    & -2.21   & 0.20    & -4.35                  & 1.18    & 10.08\tablenotemark{a} & 1.56    & 3750    & 890     \\
Tr 15-19    & -3.11   & 1.28    & -1.39   & 0.25    & -1.72                  & 1.30    & 12.71\tablenotemark{a} & 2.33    & 2630    & 690     \\
Tr 15-21    & -2.50   & 0.87    & -1.27   & 0.07    & -1.23                  & 0.87    & 13.13\tablenotemark{a} & 2.38    & 2490    & 440     \\
Tr 15-26    & -5.78   & 1.33    & -2.28   & 0.21    & -3.50                  & 1.34    & 10.70\tablenotemark{a} & 1.43    & 3590    & 970     \\
Tr 15-9     & -3.55   & 0.95    & -1.58   & 0.15    & -1.97                  & 0.96    & 12.59\tablenotemark{a} & 2.17    & 3000    & 580     \\
Tr 16-100   & \nodata & \nodata & \nodata & \nodata & -5.00\tablenotemark{b} & 0.30\tablenotemark{b}  & 8.52                   & 2.18    & 1850    & 370     \\
Tr 16-11    & -5.63   & 2.44    & -2.21   & 0.35    & -3.42                  & 2.47    & 11.20                  & 1.61    & 4000    & 1970    \\
Tr 16-115   & -7.99   & 1.08    & -2.87   & 0.06    & -5.11                  & 1.08    & 10.03                  & 1.94    & 4370    & 950     \\
Tr 16-12    & -6.03   & 1.14    & -2.84   & 0.14    & -3.19                  & 1.15    & 11.50                  & 2.26    & 3060    & 710     \\
Tr 16-122   & -4.53   & 1.12    & -2.23   & 0.23    & -2.31                  & 1.15    & 11.34                  & 1.75    & 2400    & 550     \\
Tr 16-124   & -6.33   & 1.81    & -2.84   & 0.13    & -3.49                  & 1.81    & 11.09                  & 2.24    & 2940    & 1060    \\
Tr 16-127   & -7.47   & 1.16    & -2.86   & 0.06    & -4.61                  & 1.16    & 10.67                  & 2.77    & 3170    & 740     \\
Tr 16-16    & -5.15   & 4.77    & -2.70   & 0.33    & -2.45                  & 4.78    & 10.75                  & 2.09    & 1670    & 1590    \\
Tr 16-17    & -6.69   & 2.10    & -2.72   & 0.38    & -3.98                  & 2.13    & 10.86                  & 1.83    & 4000    & 1710    \\
Tr 16-18    & -5.62   & 1.08    & -2.84   & 0.14    & -2.78                  & 1.09    & 12.11                  & 1.65    & 4440    & 970     \\
Tr 16-22    & \nodata & \nodata & \nodata & \nodata & -4.30\tablenotemark{b} & 0.30\tablenotemark{b}  & 10.85                  & 3.51    & 2130    & 430     \\
Tr 16-23    & \nodata & \nodata & \nodata & \nodata & -4.65\tablenotemark{b} & 0.30\tablenotemark{b}  & 10.00                  & 2.91    & 2230    & 450     \\
Tr 16-24    & -4.29   & 1.12    & -2.14   & 0.19    & -2.16                  & 1.14    & 11.51                  & 1.72    & 2450    & 560     \\
Tr 16-246   & -6.49   & 1.99    & -2.71   & 0.45    & -3.78                  & 2.04    & 11.92                  & 2.98    & 3510    & 1430    \\
Tr 16-26    & -7.11   & 2.57    & -2.39   & 0.45    & -4.72                  & 2.61    & 11.79                  & 2.17    & 7390    & 3870    \\
Tr 16-28    & -5.21   & 1.03    & -2.81   & 0.09    & -2.39                  & 1.03    & 11.57                  & 1.93    & 2550    & 530     \\
Tr 16-3     & -7.19   & 0.98    & -2.86   & 0.08    & -4.33                  & 0.98    & 10.12                  & 1.89    & 3250    & 640     \\
Tr 16-31    & -6.79   & 0.97    & -2.85   & 0.10    & -3.94                  & 0.97    & 10.47                  & 1.89    & 3190    & 620     \\
Tr 16-4     & -6.05   & 1.23    & -2.86   & 0.09    & -3.18                  & 1.24    & 11.17                  & 1.79    & 3260    & 810     \\
Tr 16-55    & -4.87   & 1.25    & -2.41   & 0.25    & -2.47                  & 1.28    & 12.19                  & 1.97    & 3450    & 880     \\
Tr 16-74    & -5.70   & 1.11    & -2.82   & 0.19    & -2.87                  & 1.13    & 11.61                  & 2.14    & 2940    & 670     \\
Tr 16-94    & -4.53   & 0.93    & -2.23   & 0.14    & -2.31                  & 0.94    & 9.91                   & 1.30    & 1530    & 290    \enddata
\end{deluxetable}
\tablenotetext{a}{WEBDA}
\tablenotetext{b}{Estimated based on spectral type following \citet{Walborn1972}. }

\end{document}